\documentclass[aps,prl,twocolumn,superscriptaddress,footinbib,showkeys]{revtex4-2}
\usepackage{graphicx}
\usepackage{indentfirst}
\usepackage{physics}
\usepackage{braket}
\usepackage{float}
\usepackage{amsmath}
\usepackage{epstopdf}
\usepackage{esint}
\usepackage{color}
\usepackage[T1]{fontenc}
\usepackage{subfigure}
\usepackage{amsfonts}
\usepackage{footmisc}
\usepackage{scrextend}
\usepackage{xcolor}
\usepackage{multirow}
\usepackage[english]{babel}
\usepackage{url}
\usepackage{todonotes}
\usepackage{makecell}
\usepackage[hyperfootnotes=false]{hyperref}
\definecolor{darkblue}{rgb}{0,0,0.5}
\hypersetup{
colorlinks=true,
linkcolor=black,
filecolor=blue,
citecolor=darkblue,
urlcolor=black,
}
\usepackage{cleveref}
\usepackage{mathtools}

\newcommand{\calN}{{\cal N}}
\newcommand{\calT}{{\cal T}}

\newcommand{\calS}{{\cal S}}

\def\be{\begin{equation}}
\def\ee{\end{equation}}
\def\ba{\begin{eqnarray}}
\def\ea{\end{eqnarray}}
\usepackage{bm}

\begin{document}

\title{Absolutely stable time crystals at finite temperature}

\author{Francisco Machado}
\thanks{These two authors contributed equally}
\affiliation{ITAMP, Harvard-Smithsonian Center for Astrophysics, Cambridge, Massachusetts, 02138, USA}
\affiliation{Department of Physics, Harvard University, MA 02138, USA}
\affiliation{Department of Physics, University of California, Berkeley, CA 94720, USA}
\affiliation{Materials Science Division, Lawrence Berkeley National Laboratory, Berkeley, CA 94720, USA}
\author{Quntao Zhuang}
\thanks{These two authors contributed equally}
\affiliation{Department of Physics, University of California, Berkeley, CA 94720, USA}
\affiliation{James C. Wyant College of Optical Sciences, University of Arizona, Tucson, Arizona 85721, USA}
\affiliation{Ming Hsieh Department of Electrical and Computer Engineering \& Department of Physics and Astronomy, University of Southern California, Los Angeles, CA 90089, USA}
%\email{zhuangquntao@gmail.com}
\author{Norman Y. Yao}
\affiliation{Department of Physics, Harvard University, MA 02138, USA}
\affiliation{Department of Physics, University of California, Berkeley, CA 94720, USA}
\affiliation{Materials Science Division, Lawrence Berkeley National Laboratory, Berkeley, CA 94720, USA}
%\email{norman.yao@berkeley.edu}
\author{Michael P. Zaletel}
\affiliation{Department of Physics, University of California, Berkeley, CA 94720, USA}
\affiliation{Materials Science Division, Lawrence Berkeley National Laboratory, Berkeley, CA 94720, USA}
%\email{mikezaletel@berkeley.edu}

%\date{\today}

\begin{abstract} 

      We show that locally-interacting, periodically-driven (Floquet) Hamiltonian dynamics coupled to a Langevin bath support finite-temperature discrete time crystals (DTC) with an infinite auto-correlation time. 
      By contrast to both prethermal and many-body localized DTCs, the time crystalline order we uncover is stable to arbitrary perturbations, including those that break the time translation symmetry of the underlying drive.
      Our approach utilizes a general mapping from probabilistic cellular automata (PCA) to open classical  Floquet systems undergoing continuous-time Langevin dynamics. 
Applying this mapping to a variant of the Toom cellular automata, which we dub the ``$\pi$-Toom time crystal'', leads to a 2D Floquet Hamiltonian with a finite-temperature DTC phase transition. 
We provide numerical evidence for the existence of this transition, and analyze the statistics of the finite temperature fluctuations. 
Finally, we  discuss how general results from the field of probabilistic cellular automata  imply the existence of discrete time crystals (with an infinite auto-correlation time)  in all dimensions, $d \geq 1$.

\end{abstract} 
%\keywords{Phase transitions, statistical physics, cellular automata.}
\maketitle

A tremendous amount of recent excitement has centered upon interacting periodically-driven (Floquet) ``phases of matter''~\cite{potter2016classification,khemani2016phase,else2016floquet,yao2017discrete,zhang2017observation,choi2017observation,else2020discrete}. 
While discussed as  non-equilibrium phases,  thus far attention has largely focused on two scenarios which are non-equilibrium only in a rather  restricted sense. First, there are  quantum ``many-body-localized'' (MBL) Floquet phases~\cite{nandkishore2015many,ponte2015many,khemani2016phase,else2016floquet,yao2017discrete,abanin2019colloquium,kjall2014many}. Because the ergodicity breaking of  MBL is sufficient to prevent  the periodic drive from heating the system to infinite temperature, the system does not need to be coupled to a dissipative bath (e.g., the dynamics are driven, but purely unitary)~\cite{bukov2015universal,abanin2016theory,weidinger2017floquet}. 
In this case, the eigenstates of the Floquet evolution have area-law entanglement, which allows much of the physics to be mapped to more familiar questions of order in quantum ground states~\cite{huse2013localization,chandran2014many,bahri2015localization,potirniche2017floquet}.
Second, there are  ``prethermal'' Floquet phases, in both classical and quantum systems, which heat  exponentially slowly due to (for example) a mismatch between the driving frequency and the natural frequencies of the undriven system~\cite{abanin2015exponentially,else2017prethermal,zeng2017prethermal,mori2018floquet,machado2019exponentially,machado2020long,kyprianidis2021observation,Ye2021Floquet,pizzi2021classical,else2020long}. During the exponentially long time-scale before heating, these systems can exhibit behavior which is  analogous to order in finite temperature equilibrium phases~\cite{else2017prethermal,machado2020long,Ye2021Floquet,pizzi2021classical}.
However,  prethermal Floquet phases  are not ``true'' phases in the strict sense because they are distinguished from disordered behavior via crossovers, rather than sharp transitions~\cite{else2020discrete}.
% \todo[inline]{Perhaps we can simplify this a bit, removing the prethermal stuff [  :(  ] and focusing on MBL and ergodicity breaking}

Perhaps the most paradigmatic example of a Floquet phase of matter is the so-called discrete time-crystal (DTC)---starting from a generic initial state, at long times the DTC  relaxes into a steady state with a temporal periodicity which is a multiple of the  drive's~\cite{khemani2016phase,else2016floquet,yao2017discrete}. 
This behavior breaks the discrete time-translation symmetry of the drive and is stable to small  perturbations of the dynamics.
\begin{figure}[H]
\centering
\includegraphics[width=3.0in]{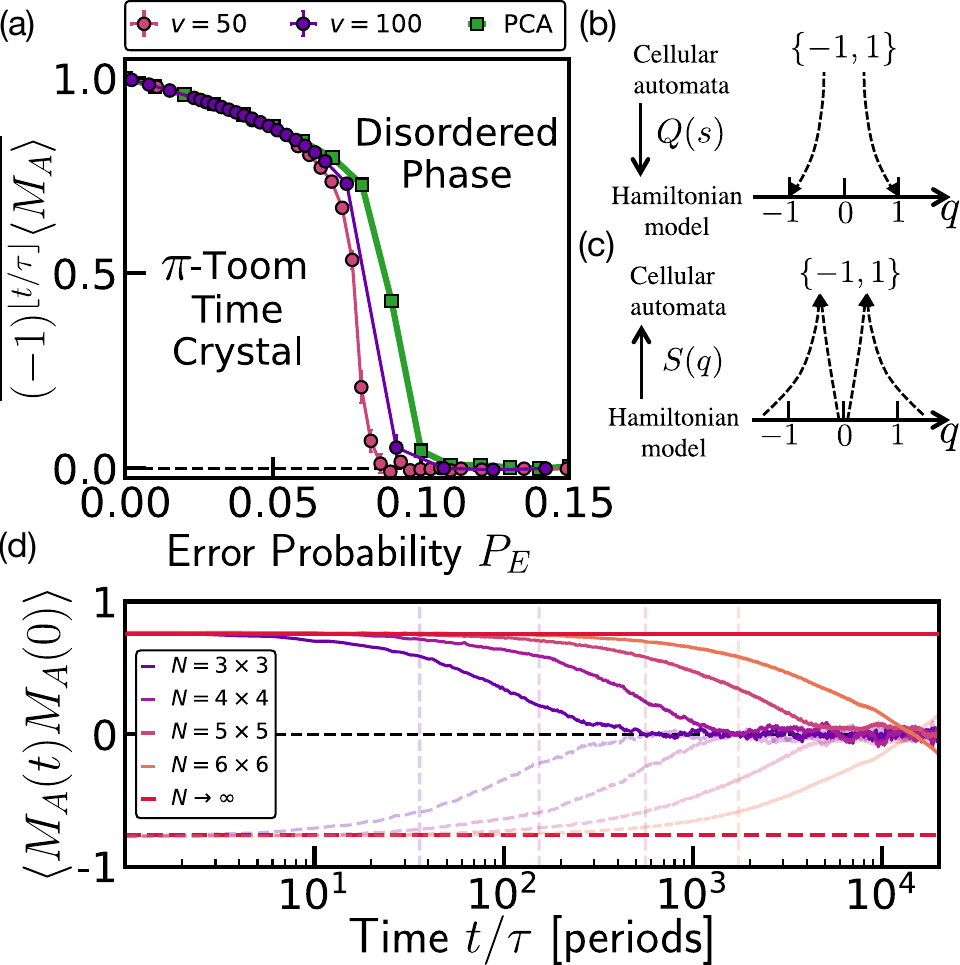}
\caption{(a) Time crystalline order parameter (e.g.~stroboscopic magnetization) as a function of the error probability. 
The phase transition from a discrete time crystal to the disordered phase is shown for both a 
continuous-time, Floquet Langevin simulation of the $\pi$-Toom model  with pinning potential $v=50,100$, as well as for a direct implementation of the $\pi$-Toom PCA. 
We average $(-1)^{\lfloor t/\tau \rfloor} \braket{M_A}$ starting at $t=3000$ for $\sim 500$ Floquet cycles, $\sim 50$ noise realizations and system size $N=32\times32$.
(b,c) Schematic of the translation between the discrete state space of a cellular automata and the continuous state space of a Hamiltonian model. 
(d) Dynamics of the Langevin $\pi$-Toom model ($v=100, T=7$), exhibiting robust period doubling (full line even times, light dashed line odd times) with a lifetime that grows exponentially with $L$~\cite{supp}. 
We infer that for the system size ($L=32$) considered in panel (a), the lifetime of the $\pi$-Toom time crystal is significantly longer that the time window over which the magnetization is averaged. 
}
\label{Hmodel_schematic_state}
\end{figure}

% \noindent 
%

%Similar behavior has been studied in the non-linear dynamics community as either ``subharmonic entrainment'' or ``asymptotic periodicity''~\cite{lasota1984asymptotic,Bennett1990Stability,losson1995phase,losson1996thermodynamic,gielis2000coupled,parlitz1997subharmonic}; for a discussion of how this earlier work differs from results in the quantum many-body context, we refer the reader to~\cite{yao2020classical,else2020discrete,khemani2019brief}. 

%From the perspective of symmetry breaking, it is surprising that ``true'' time crystalline order has only been settled in MBL systems (i.e. where the lifetime of the DTC order diverges exponentially in the system size).

%However, without a mechanism to prevent heating and break ergodicity in 

From the perspective of  symmetry breaking, it is natural to ask if ``true'' time crystals, with an infinite autocorrelation time exist beyond many-body localized quantum systems~\cite{zaletel2023colloquium}; 
here, we emphasize that the lifetime of the DTC order should diverge exponentially in the system size as the thermodynamic limit is taken, while all other parameters (i.e.~Floquet frequency, temperature, etc.) are kept fixed.
% \todo[inline]{Maybe here we can just emphasize: not MBL as in there are no conserved quantities (emergent or not)}
%
Without MBL to prevent heating, one requires an alternate strategy to stabilize time crystalline order; one approach that is compatible with both quantum and classical many-body dynamics is to couple the system to a dissipative bath~\cite{else2017prethermal,tucker2018shattered,lledo2019driven,buca:2019a,cosme2019time,gambetta:2019,gambetta:2019a,lazarides:2020,dogra:2019,kessler2020continuous,kessler:2021,taheri:2021}.

%
%Combining periodic driving and a dissipative bath introduces the full complexity of non-equilibrium dynamics~\cite{lazarides2017fate,tucker2018shattered,lledo2019driven,buca:2019a,cosme2019time,gambetta:2019,gambetta:2019a,lazarides:2020,dogra:2019,kessler2020continuous,kessler:2021,taheri:2021}.   

%\todo[inline]{In fact }

%\add{While recent theoretical and analytical work offers important insights into the wealth of behavior that arise in this context \cite{buca:2019a,gambetta:2019,gambetta:2019a,lazarides:2020,dogra:2019,kessler:2021,taheri:2021}, we still lack a complete understanding of the requirements for the absolute stability of the DTC phase.}
Thinking microscopically, classical driven dissipative systems are described by Hamiltonian dynamics coupled to a finite-temperature Langevin bath, or in the quantum case, periodically-driven Lindblad evolution.
A key feature in both these contexts is that if the bath is dissipative, at finite temperature it should also come with noise due to the fluctuation-dissipation theorem. At zero temperature, where there is damping but no noise, many-body time-crystals can occur rather trivially by analogy to the ``period doubling'' of coupled iterated logistic maps~\cite{kaneko1984period,kapral1985pattern,bunimovich1988spacetime,kaneko1992overview,kaneko1987transition}.
By contrast, the presence of noise pushes the system to explore its entire phase space and can therefore desynchronize any initial time-crystalline behavior.
This noise-induced ergodicity leads to a finite lifetime for the DTC order. 
The key question we will focus on is the following: Can true time-crystals exist in a periodically driven system of locally interacting particles coupled to an equilibrium bath at finite temperature?

In this Letter, we argue in the affirmative:  finite-temperature time-crystals~\cite{wilczek2012quantum,shapere2012classical}, with an \emph{infinite} auto-correlation time, can exist even in translation-invariant arrays of classical non-linear oscillators interacting only with their nearest-neighbors. 
    This significantly strengthens and extends upon prior works which found that arrays of parametrically-driven oscillators did not produce a stable time crystal: thermal fluctuations led to an auto-correlation time which was finite and activated~\cite{yao2020classical}.

    Our results are three-fold. First, we explain~\cite{Bennett1990Stability} how  non-trivial results in the field of probabilistic cellular automata (PCA)~\cite{von1968general,wolfram1983statistical} imply that local PCAs can exhibit  time-crystalline behavior stable to arbitrary small perturbations in any dimension, $d \ge 1$~\cite{gacs2001reliable,gray2001reader,toom1980stable,bennett1985role,makowiec1999stationary}.
    %
%A PCA is a deterministic cellular automata (CA) perturbed by stochastic errors, mimicking the effect of finite temperature, making it a ``stripped-down'' model of  finite-temperature non-equilibrium dynamics.
%
%We first use the results of G{\'a}cs 
%
Unlike MBL or prethermal time crystals, such  time-crystalline order is ``absolutely stable'', in the sense that it remains robust \emph{even} in the presence of perturbations that break the discrete time-translation symmetry of the periodic drive. 
Next, we extend these PCA results to the physical setting of interest---classical, continuous-time, Langevin dynamics.
% ---  which is constrained, for example, by symplectic structure and the fluctuation-dissipation theorem. 
%
In particular, we provide an explicit protocol which allows classical Langevin dynamics  to ``simulate'' any PCA;  focusing on the example of the so-called $\pi$-Toom PCA in 2D~\cite{toom1980stable,bennett1985role,makowiec1999stationary}, we find that our continuous-time Langevin simulation indeed exhibits a finite-temperature phase transition between a discrete time-crystal and a disordered phase (Fig.~\ref{Hmodel_schematic_state}).  
Finally, we utilize results from  ``large deviations'' theory to analytically obtain a bound on the spatio-temporal error cumulants; building on this bound, we perform an extensive analysis of the errors due to Langevin noise and provide numerical evidence that they are of a type covered by rigorous mathematical results from G{\'a}cs  and Toom~\cite{gacs2001reliable,gray2001reader,toom1980stable,bennett1985role,makowiec1999stationary}.

%
%Applying this to a 2D array of locally interacting mechanical oscillators, continuous-time Langevin simulations (Fig.~\ref{Hmodel_schematic_state}) reveal a  finite-temperature phase transition between a discrete time-crystal and a disordered phase. 
   
   %Our work thus shows that discrete time crystals can exist in driven classical systems coupled to a bath, even when the periodic drive is imperfect. 
   %In contrast to  previous mechanisms for realizing DTC order, including many-body localization and prethermalization, the perturbations can even systematically violate the time-translation symmetry of the drive, making such behavior ``absolutely stable'' in the strongest sense. 
   
   %
   %Finally, 

\emph{Time-crystals in probabilistic cellular automata}---The state of a probabilistic cellular automata is given by a spin configuration $\{\eta(\bm{x})\}$, where $\bm{x} \in \Lambda$ labels sites in a regular lattice $\Lambda$ and each $\eta(\bm{x})$ takes values in a finite set $\mathcal{S}$.
In a conventional cellular automata (CA), the dynamics are governed by a deterministic transition rule~\cite{ulam1952random,neumann1966theory}, 
\begin{align}
\{ \eta(\bm{x}, t+1) \} &= \mathcal{T}[ \{\eta(\bm{x}, t) \}].
\end{align}
In a PCA, the spins instead evolve under a Markov process described by the transition matrix $M_{ \eta  \to   \eta' }$, which characterizes  the \emph{probability} to evolve from configuration $\eta$ to $\eta'$~\cite{toom1974nonergodic,dawson1977stable,gacs1986reliable}. 
%$M$ should be local in the sense that the update  distribution of a spin  depends only on the state of its nearest neighbors, or more generally, on some finite range ``neighborhood'' $ \mathcal{N}$. 
    %
%A particularly natural class of PCAs arise by starting with a deterministic CA and perturbing it with an ``error rate'' $\epsilon$.
%More concretely, at each step, the spins first follow the rule $\mathcal{T}$, and then with a probability bounded by $\epsilon$ they randomly flip  to a different state. One can think of the resulting Markov process as a perturbation to the deterministic one, $M = \mathcal{T} + \epsilon \, \Delta M$, where $\Delta M$ determines the precise error distribution. The mathematical results we will describe can in fact account for even more general (non-Markovian) error models, as we will describe shortly~\cite{gacs2001reliable, gacs2021new}. 

%A deterministic CA (with $\calS=\{-1,1\}$) can trivially realize a time-crystal:  for example, the rule $1 \leftrightarrow -1$. In fact, since CA are capable of  universal classical computation, they can realize any dynamical phenomena which can be programmed on a computer~\cite{codd2014cellular}. 

Whether a PCA can realize a stable time-crystal is subtle. 
The long-time dynamics of a PCA are described by the stationary probability distributions $\mathcal{P}[\eta]$ of $M$, e.g.,  $M \mathcal{P}[\eta] = \mathcal{P}[\eta]$.
We say $M$ exhibits an $n$-fold subharmonic response if there are $n > 1$ distinct distributions, $\mathcal{P}_i[\eta]$, such that $M \mathcal{P}_i[\eta] = \mathcal{P}_{i+1}[\eta]$, with $\mathcal{P}_{n} = \mathcal{P}_{0}$.
This simply formalizes the notion of long-time oscillations: at long times, a generic initial state will relax into to a non-uniform convex combination $\sum_i p_i \mathcal{P}_i$ which is stationary under $M^n$, but not $M$~\cite{footnote1}.
A PCA time crystal is then defined to be a local PCA with a \emph{stable}  $n$-fold subharmonic response.
This motivates our first question: Do PCA time crystals exist~\cite{footnote2}?

One prerequisite for a PCA time-crystal is ergodicity breaking. A PCA is ``ergodic'' if it has a unique stationary distribution, so that at long times the state is independent of the initial spin configuration.
A time-crystal necessarily breaks ergodicity  because $M^n$ has $n$ stationary distributions, so the system remembers which of the $n$ states in the orbit it is in.
Ergodicity-breaking PCAs were first proved to exist in 2D  by Toom~\cite{toom1974nonergodic,toom1980stable}, and much later in 1D by G{\'a}cs~\cite{gacs2001reliable}. 
We focus our discussions here on Toom's model  because of its simplicity (a discussion of the G{\'a}cs model in 1D is provided in the supplemental material~\cite{supp}).  
The Toom model is a 2D PCA with a binary state space, $\calS=\{-1,1\}$, and a ``majority vote'' transition rule in the Northern-Eastern-Center (NEC) neighborhood $\calN=\{(1,0),(0,1),(0,0)\}$  [the $(\Delta x_i, \Delta y_i) \in \mathcal{N}$ denote the relative locations of the cells in the neighborhood].

Crucially, in this model, it was proven that there are two ``phases'' (i.e.~stationary distributions), corresponding to states ``all $+1$'' and ``all $-1$'', which are stable against arbitrary stochastic perturbations below a critical error rate $\epsilon$~\cite{fn9}, provided that the errors are not too correlated~\cite{toom1980stable, gacs2001reliable, gacs2021new}.
This requirement can be encoded into an \emph{error condition}: there exists $\epsilon$ such that the probability of $k$ errors is bounded:
%encoded in the following :
\begin{align}
P_{\wedge_{\ell=1}^k E_{u_\ell}}=\prod_{\ell=1}^k P_{E_{u_\ell}|E_{u_{\ell-1}}...E_{u_1}}\le \epsilon^k. 
\label{error_condition}
\end{align}
where $E_u$ denotes an ``error'' at a space-time point $u$, i.e. the dynamics did not follow the original update rule $\mathcal{T}$.

Finally, a simple modification of the Toom PCA~\cite{Bennett1990Stability,gielis2000coupled}, which we call the ``$\pi$-Toom'' model, immediately turns his construction into a time-crystal exhibiting an infinite autocorrelation time~\cite{fn3}: in particular, instead of an NEC majority vote, one utilizes an NEC anti-majority vote, or equivalently, one  considers the Toom model with an interleaved spin flip, $1 \leftrightarrow -1$, between each step. 
In 1D, an equivalent construction based upon the G{\'a}cs model yields a similar PCA time crystal.
%

%% As we will later explore in detail, G{\'a}cs' and Toom's results depend upon important assumptions  about the error model~\cite{toom1980stable, gacs2001reliable, gacs2021new}.
%% %
%% Since the PCA can naturally be viewed as a perturbation to a cellular automata rule $\mathcal{T}$,  given a particular spatio-temporal history $\eta(\bm{x}, t)$, one can denote an ``error'' $E_u$ having occurred  at space-time point $u$, if it did not follow the original update rule.
%% %; instead, the state of $u$ is chosen according to some noise distribution.
%% %
%% % is only correct?
%% The only requirement is that the probability of errors occurring at $k$ space-time points, $\{u_{\ell} \}$,  satisfies the  bound:
%% \begin{align}
%% P_{\wedge_{\ell=1}^k E_{u_\ell}}=\prod_{\ell=1}^k P_{E_\ell|E_{\ell-1}\cdots E_1}\le \epsilon^k.
%% \label{error_condition}
%% \end{align}
%% for some constant $\epsilon$ \cite{gacs2001reliable}.

\emph{Simulating a PCA using Floquet-Langevin  dynamics}---In order to extend PCA time crystals to a microscopic, physical setting, we now demonstrate the ability for continuous-time Langevin dynamics to effectively simulate \emph{any} PCA.
The dynamics we consider take the general Langevin form:
\begin{align} \label{Langevin_dynamics}
\dot{q}_i &= \partial_{p_i} H( \{p, q \}; t) \notag \\[3pt]
\dot{p}_i &= -\partial_{q_i} H(\{p, q \}; t) + R_i(t) - \gamma p_i \\[3pt]
\langle R_i(t)& R_j(t') \rangle_{\textrm{noise}} = 2 \gamma T \delta_{ij} \delta(t - t') \notag ,
\end{align}
\noindent where $(q_i, p_i)$ are the conjugate variables of a classical oscillator  at site $i$, and $R_i(t)$ is a stochastic force whose variance is proportional to the friction coefficient $\gamma$ and the temperature $T$.
%
%As depicted in Fig.~\ref{Hmodel_schematic_state}(a,b), we will encode the discrete state of the cellular automata spin, $\eta_i$,  as the integer part of the position, $\eta_i =  \floor*{q_i}$~\cite{fn8}.
%
We focus on a CA with states $\eta = -1, 1$ which we will encode in the oscillator with the identification $\eta = \mathrm{sign} (q)$ as depicted in Fig.~\ref{Hmodel_schematic_state}(b,c).

To be specific, we focus on $\eta = \pm 1$ and we will encode $\eta$ as the sign of the oscillator position $q$.
The Hamiltonian  takes the  form $H(t) = \sum_i \frac{p^2_i}{2 m} + U(\{q\}, t)$, with $U(t)$  engineered so that one Floquet cycle, $H(t + \tau) = H(t)$, will implement  one  application of the  update rule $\mathcal{T}$.

When attempting to build Floquet-Langevin dynamics that simulate a PCA, we encounter the following challenge.
In our continuous-time dynamics, one needs a way to ``store'' the previous global state throughout the update cycle.   
This is essential in order to give the  dynamics enough time to identify what the new state of the system should be. 
To solve this issue, at each position $\bm{x}$, we envision \emph{two} oscillators ($A$ and $B$) with coordinates $(q_{\bm{x}}^A,p^A_{\bm{x}})$ and $(q_{\bm{x}}^B,p^B_{\bm{x}})$.
At each step, we will view one set of oscillators (say $A$) as the ``memory'', while the other set ($B$) will undergo evolution to the new state $B = \mathcal{T}(A)$, driven by $U(\{q\}, t) = V_{\mathrm{I}}(\{q\})$.
After letting the particles relax to their positions using a pinning potential $U(\{q\}, t) = \sum_i V_{\mathrm{pin}}(q_i)$, we exchange the role of $A$ and $B$ and repeat. 
For more details on the specific four-step Floquet-Langevin sequence see the appendix.

\begin{figure}[t]
\includegraphics[width=3.4in]{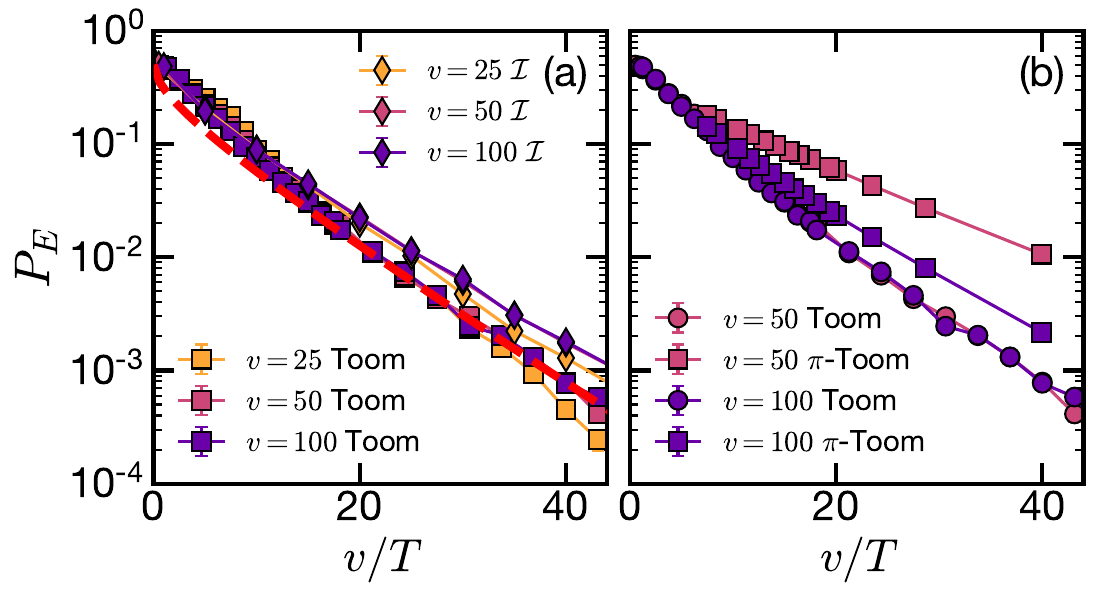}
\caption{
  Error probability $P_E$ versus the ratio of the pinning potential to the temperature, $v/T$, in simulations of (\textbf{a}) the do-nothing  ($\mathcal{I}$) and Toom CAs.
  The dashed-red line indicates the equilibrium estimate $P_E = \frac{1}{2}{\rm Erfc}\left (\sqrt{v_I/(2T)} \right)$. (\textbf{b}) The $\pi$-Toom CA.
While $P_E$ apparently depends on the simulated CA, as $v$ increases the $\pi$-Toom error rate converges toward the  Toom error rate. In all cases, we find an  exponential decay in the error rate as a function of $v/T$.
Data are obtained from a $32\times 32$ system by averaging  over 25 Floquet cycles after an initial evolution of 200 Floquet cycles.
\label{PE_benchmark}}
\end{figure}

\emph{Discrete time-crystal in a Floquet-Langevin simulation of the $\pi$-Toom model}---Within the PCA setting, the $\pi$-Toom model is a discrete time-crystal, and we have just described a procedure for ``simulating'' any PCA using continuous-time Floquet-Langevin dynamics. 
Naively, it seems that combining these two insights immediately yields a continuous-time, Floquet-Langevin DTC with an infinite lifetime. 
The subtlety is the following:  at finite temperature, the errors due to Langevin noise (e.g.~thermally activated escape out of the pinning potentials) may not satisfy the requirements of the G{\'a}cs and Toom error models [Eq.~\ref{error_condition}].
We will return to a detailed analysis of error correlations, but let us begin by numerically exploring the existence of time crystalline order in a Floquet-Langevin simulation of the $\pi$-Toom model. 

Working with a two dimensional square lattice, we perform an extensive set of  numerical simulations by  solving the Floquet-Langevin dynamics [Eq.~\eqref{Langevin_dynamics}] using a symplectic stepping method. 
To implement the $\pi$-Toom model, we take the pinning potential to be: 
$V_{\textrm{pin}}(q)= v_{\textrm{pin}} (q-1)^2 (q+1)^2+ F q$, where $F=10^{-4}$ breaks the  accidental Ising symmetry of the  model. %
We parameterize the magnitude of the interaction and the pinning potentials as $4v_I=v_{\textrm{pin}}=v$, while the noise term, $R_i(t)$, is implemented via random momentum kicks  with variance $2\gamma T dt$.
%, where $dt$ is chosen to ensure that the relative momentum change within each step is  small. 
%
Finally,  $\gamma$ is chosen such that the dynamics are tuned to critical damping relative to both $V_{\textrm{pin}}$ and $V_{I}$.

In order to ensure that a single Floquet period implements only one $\pi$-Toom update, we utilize the following interleaving strategy: in step two, we choose $\mathcal{T}$ to be the Toom update rule, while in step four, we choose $\mathcal{T}$ to be the $\pi$-Toom update rule. 
We compute the  Floquet-Langevin dynamics up to time-scale, $t \sim 10^4$, starting from a uniform initial state (a study of different initial states is provided in the supplemental material~\cite{supp}). 
We probe the resulting dynamics by measuring the  average ``magnetization'', $\braket{M_A}\equiv  \frac{1}{N} \sum_k {\rm sign}(q_k^A)$, where $N = L \times L$ is the system size. 
Time crystalline order corresponds to stable period-doubling of the magnetization; indeed, as shown in Fig.~\ref{Hmodel_schematic_state}(d), the auto-correlation of the magnetization $\langle M_A(t) M_A(0)\rangle$ exhibits period-doubling out to a time-scale that increases exponentially with $L$ \cite{supp}.
The DTC order parameter is then defined as the time average (over both space \emph{and} time) of the stroboscopic magnetization: $\overline{(-1)^{\lfloor t/\tau \rfloor} \braket{M_A}}$.

In order to compare our Floquet Langevin simulation with a direct implementation of the $\pi$-Toom PCA, we first translate the temperature, $T$, to an effective error rate $P_E$ (per space-time unit cell)~\cite{fn7}. 
As shown in Fig.~\ref{Hmodel_schematic_state}(a), the time crystalline order parameter, which we estimate from the time-window $t/\tau = 3000 - 3500$,  
exhibits a  phase transition as a function of $P_E$.
The functional form and location of the DTC phase transition are in good agreement between our continuous-time Floquet Langevin simulation of the $\pi$-Toom model and a direct implementation of the PCA itself (with improving agreement for larger pinning potentials).

\emph{The nature of errors in the Floquet-Langevin DTC}---Due to the presence of a finite temperature bath, our continuous-time Floquet-Langevin simulation of the $\pi$-Toom model is intrinsically noisy. 
Large  thermal fluctuations can lead to an ``error'' in the subsequent state  $\eta({\bm x},t)$ relative to the  noiseless transition $\calT\big(\eta({\bm x}+\calN,t-1)\big)$.
Fortunately, our overall goal is in fact to simulate the noisy PCA version of the $\pi$-Toom model. 
However, even then, the  distribution of errors arising from the Floquet-Langevin dynamics need not (a priori) be consistent with the aforementioned error condition [Eq.~\ref{error_condition}].

\begin{figure}
\centering
\includegraphics[width=3.4in]{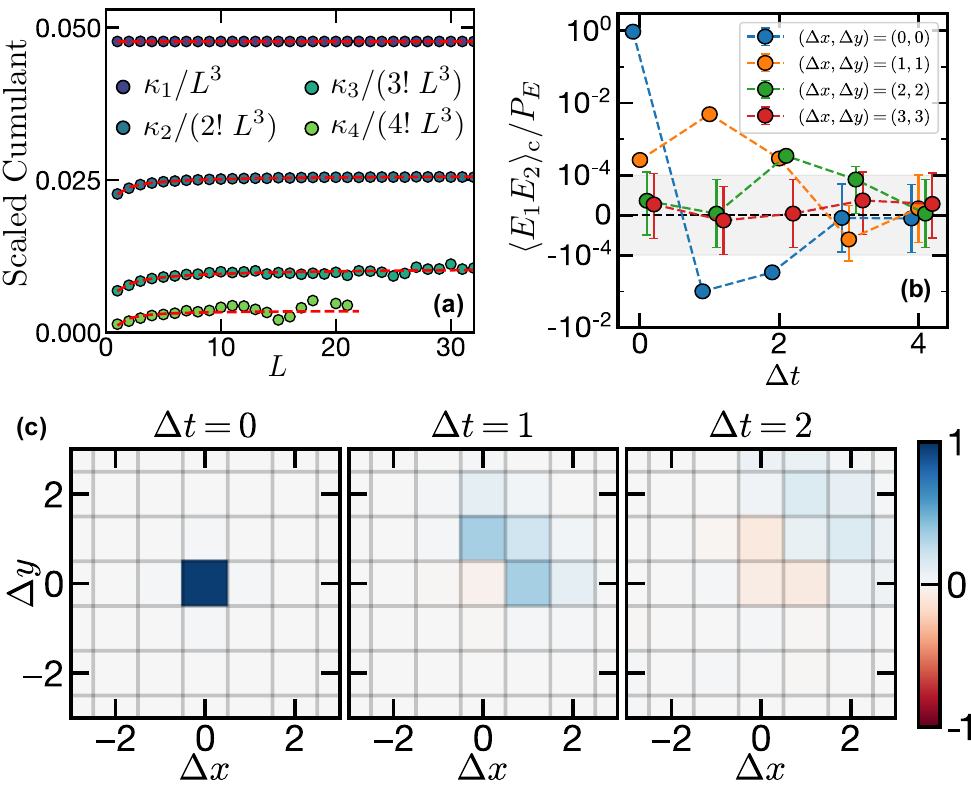}
\caption{(a) Scaled cumulants, $\kappa_n/(L^3n!)$, of the error distribution from a Floquet Langevin simulation of the $\pi$-Toom model for $L \times L\times L$ space-time volumes. The dashed curves are fits to $\kappa_n / L^3  = c_n  - b_n L^{-\mu_n}$. 
As shown in the supplemental material~\cite{supp}, we find $\mu > 0$ indicating convergence to a finite $c_n$. 
Each data point was estimated from the statistics of $3000$ independent Langevin trajectories ($T=5.17, v=100$), with $1000$ $L \times L\times L$ blocks  sampled from each trajectory.
(b,c) Shows the connected two-point correlations of the errors for $v=100$ and $T=5.17$. The 
magnitude of the error correlations for a diagonal cut along space are depicted in (b), while the real-space propagation of errors is shown in (c). The color map range is rescaled by 1.0, 0.025, 0.004 for the left, middle and right panels respectively.
}
 \label{fig:piToom_Error_Corr}
\end{figure}

To this end, our final goal is to obtain numerical evidence that:~(i) the errors arising from the Floquet-Langevin dynamics satisfy Eq.~\ref{error_condition} for some constant $\epsilon(T)$ and (ii) the error bound $\epsilon(T)$ can be made arbitrarily small as  $T \to 0$.
To begin, we first examine the temperature dependence of the error \emph{rate} per space-time unit cell $P_E = \langle E_u \rangle$, where $E_u = 0/1$ is the indicator function for an error at $u$~\cite{fn10}.
%, and show  that $P_E(T)$ decays exponentially as $T \to 0$.
In Fig.~\ref{PE_benchmark}, we show the empirically measured error rate $P_E(T)$ as a function of $v/T$ for the Floquet-Langevin simulation of three different  PCA rules:  the ``do-nothing'' rule $\mathcal{I}$, the Toom rule, and the $\pi$-Toom rule. 
In all cases we find that decreasing the temperature leads to an exponential decay in $P_E$,  implying that for  strong potentials and low temperatures, arbitrarily small error probabilities are obtained.

We now turn to the crucial issue of spatio-temporal correlations.
Consider an arbitrary space-time volume $V$ containing $|V|$ points. 
Letting $P(N_V)$ denote the probability that $N_V$ errors occur in the volume $V$, we aim to provide empirical evidence that there is a constant $\epsilon$ such that $P(N_V = |V|) \leq \epsilon^{|V|}$ for all $V$.
However, measuring  $P(N_V = |V|)$ directly is difficult because for large $|V|$ such ``large deviations''~\cite{deuschel2001large,varadhan1984large,den2008large,touchette2011basic} are too rare to enable the collection of sufficient statistics. 
To make progress, we will instead relate $P(N_V = |V|)$ to the connected $n$-point correlations of the errors. %, which are feasible to estimate for low enough $n$. 
%
%Roughly speaking, if the connected $n$-point functions decay fast enough, our desired error-bound will be satisfied.
%
In particular, following the detailed derivations in appendix, our aim is to provide numerical evidence that the $n^{\textrm{th}}$ cumulant, $\kappa_n$ does not grow faster than $n!$ or $|V|$.

Restricting to space-time boxes of dimension $|V| = L^3$, we show the estimated  cumulants for our Floquet-Langevin simulation of the $\pi$-Toom model for $2 \le L \le 32$ [Fig.~\ref{fig:piToom_Error_Corr}(a)]. They converge to a finite $c_n$ with a power law correction in $1/L$~\cite{supp}. While it is difficult to estimate the cumulants beyond $n>3$, from the available data, the $n!$ bound on $c_n$ is safely satisfied.
In addition to power-law correlations, the non-zero higher-order cumulants also suggest that the error distribution is  non-Gaussian.
Interestingly, this originates from the anisotropic nature of the $\pi$-Toom update rule. 
An initial error  causes an increased likelihood for errors at nearby space-time points, with correlations that propagate outward in the ``NEC'' direction [Fig.~\ref{fig:piToom_Error_Corr}(b,c)].

In summary, despite errors which are both power-law correlated and non-Gaussian, we find compelling evidence that the Floquet-Langevin DTC satisfies the requisite error condition for absolute stability~\cite{gacs2001reliable,gray2001reader,toom1980stable,bennett1985role,makowiec1999stationary}. 
Of course we cannot rule out that for some anomalously large volume $|V|$,  cumulant order $n$, or  inverse temperature $1/T$, the observed behavior will change course and violate the bound---a caveat common to any numerical finite-scaling approach. 
Obtaining a rigorous proof of this bound thus remains an important open question.

\emph{Acknowledgements}---
%Q.Z.~and~F.M.~contributed equally to this work. 
We are greatly indebted to conversations with P. G\'acs for generously explaining various details of his work, C. Maes for pointing us to the literature on coupled map lattices, and C. Nayak and L. Balents for collaborations on related work. MPZ is indebted to conversations with D. Huse during the initiation of this work. This work is supported in part by the Army Research Office (Grant No. W911NF2110262), the DARPA DRINQS program (Grant No. D18AC00033), the A. P. Sloan foundation, the David and Lucile Packard foundation, and the W. M. Keck Foundation.

\emph{Note added}: During the completion of this work, we
became aware of complementary work exploring the utility of Toom-like dynamics for stabilizing time-crystalline phases in noisy, incoherent quantum spin systems~\cite{mcginley2022absolutely}.

\appendix
\renewcommand{\thefigure}{A\arabic{figure}}
\setcounter{figure}{0}

\section{Appendix on details of the specific four-step Floquet-Langevin dynamics}
\label{appendixA}

We begin with the oscillators at site $\bm{x}$ in the state $(q^{A/B}_{\bm x}, p^{A/B}_{\bm x}) = \Big(\eta(\bm{x},t \big), 0\Big)$.
From there, the dynamics evolve via a 4-step process governed by~\cite{fn6}:
\be
H\left(t\right)=\sum_{\bm{x}}\frac{{p^A_{\bm{x}}}^2}{2m}+ \frac{{p^B_{\bm{x}}}^2}{2m}+U\left(t,\{q^A_{\bm{x}}, q^B_{\bm{x}}\}\right),
\ee
where the potential $U(t, \{q^A_{\bm{x}}, q^B_{\bm{x}}\})$ has a Floquet period of $\tau=4$:
\begin{align}
&U\left(t,\{q^A_k, q^B_k\}\right)=
\\
& =\begin{cases}
  \sum_{\bm{x}} V_{\textrm{pin}}(q^A_{\bm{x}})+V_{\textrm{pin}}(q^B_{\bm{x}}) 
  & \mod(\lfloor t\rfloor,4)=0,2\\
  \sum_{\bm{x}} V_{\textrm{pin}}(q^A_{\bm{x}})+ V_I\left(q^A_{\bm{x}+\calN},q^B_{\bm{x}}\right) 
  & \mod(\lfloor t\rfloor,4)=1\\
  \sum_{\bm{x}} V_I\left(q^B_{\bm{x}+\calN},q^A_{\bm{x}}\right) + V_{\textrm{pin}}(q^B_{\bm{x}})
  & \mod(\lfloor t\rfloor,4)=3.
\end{cases} \notag
\label{Vt}
\end{align}
Let us unpack each of these steps in turn. 
First, we envision turning on a one-body potential, $V_{\textrm{pin}}(q)$, which has been engineered so that so that each $q_{\bm{x}}$ has local minima at $q = \pm 1$. 
At sufficiently low temperatures, the dissipative dynamics [Eq.~\ref{Langevin_dynamics}] will relax each oscillator's positions, $q_{\bm{x}}$, toward the nearest local minima, which corresponds to valid state of the CA.  
The second step of the Floquet dynamics implements the cellular automata transition $B = \mathcal{T}(A)$.
We will keep $q^A$ fixed using the pinning potential.
For the $B$ oscillators, however, we turn off $V_{\textrm{pin}}$, and turn on an interaction potential $V_I$ between oscillators A and B.
The interaction is engineered such that each oscillator B experiences a single potential minimum corresponding to the desired update rule as dictated by the oscillators A in its neighborhood; as a result, $V_I$  depends on both $q_{\bm{x}}^B$ and $q_{\bm{x}+\mathcal{N}}^A$.
In the third step,  we turn off the interaction, $V_I$, while ramping up the pinning potential, $V_{\textrm{pin}}$.
As in the first step, dissipation relaxes and pins the positions of the oscillators.
Finally, in the last step, we implement ``$A = \mathcal{T}(B)$'' by repeating step two with the role of $A$ and $B$ reversed~\cite{supp}.
Note that one can also replace the transition $\mathcal{T}$ in the last step with the “do-nothing” CA rule, $\mathcal{I}$, if one wants to implement only a single CA update per Floquet cycle.
%flm{Should we make a comment here that $\mathcal{T}$ need not be another implementation of the CA }
%
%After these four steps, our Floquet dynamics  have implemented two steps of the cellular automata update rule, $\cal T$.
%This block naturally forms a single period of the Floquet drive, which can then  be repeated~\cite{fn5}.
%

\section{Appendix on derivations relating $P(N_V = |V|)$ to the connected $n$-point correlations of the errors}
\label{appendixB}
%VP=VI
%Vpin=VB

Consider the scaled cumulant generating function (SCGF), $\lambda_V(j) = \frac{1}{|V|} \log  \langle e^{j N_V} \rangle$, which upper bounds the error probability as $P(N_V = |V|) \leq e^{ - |V|(j -  \lambda_V(j))}$ for any choice of $j \geq 0$.
The error bound from Eq.~\ref{error_condition} can then be re-expressed via a min-max principle as:
\begin{align}
\ln(1/\epsilon) = \min_V \max_{j \geq 0} \left[j - \lambda_V(j)\right]. 
\end{align}
Crucially, the Taylor series of the SCGF is directly related to the cumulants, $\kappa_n$,  of the error distribution, $\lambda_V(j)= |V|^{-1} \sum_{n=1}^\infty  \kappa_n \frac{j^n}{n!}$; the cumulants themselves are in turn directly related to the connected correlations, e.g.~for $n=2$,  $\kappa_2 =\sum_{u_1, u_2 \in V} \langle  E_{u_1} E_{u_2} \rangle_\textrm{c}$.
If the correlations decay sufficiently rapidly as a function of distance, the cumulants will scale as $\kappa_n \sim |V| c_n(V)$, where the coefficient $c_n(V)$ depends on the geometry of $V$ but does not grow with $|V|$ (i.e.~$c_1(V) = P_E$).
So long as this coefficient is upper bounded by some constant $c_n \equiv \max_V c_n(V)$, and $c_n$ itself grows slower than $n!$, one can immediately upper bound $\lambda_V(j) \le  \lambda(k) \equiv \sum_{n=1}^\infty c_n \frac{ k^n}{n!}$ for all $k \geq 0$.
It then follows that $\ln(1/\epsilon) = \max_{j \geq 0} [j - \lambda(j)]$ also satisfies Eq.~\eqref{error_condition}.
Since $\lambda(0) = 0$ and $\lambda'(k) = P_E < 1$, the maximal value is positive and finite, ensuring Eq.~\eqref{error_condition} is satisfied for some $\epsilon < 1$.
%
%so that we obtain a  bound $\lambda_V(k) \leq  \lambda(k) \equiv \sum_{n=1}^\infty c_n \frac{ k^n}{n!}$ for all $k \geq 0$.
%It would then follow that $\ln(1/\epsilon) = \max_{k \geq 0} (k - \lambda(k))$.
%Since $\lambda(0) = 0$ and $\lambda'(k) = P_E < 1$, the maximal value is  positive and finite, ensuring Eq.~\eqref{error_condition} is satisfied for some $\epsilon < 1$.
%
%$\kappa_n$
%

%% The above analysis can be further strengthened by relating the scaled cumulant generating function to the existence of a finite $\epsilon$ that satisfies Eq.~\ref{error_condition}. To this end, consider the SCGF $\lambda_V(k)$ defined by $\langle e^{k N_V}  \rangle = e^{|V| \lambda_V(k)}$. 
%% The SCGF upper bounds  $P(N_V = |V|) \leq e^{ - |V|(k -  \lambda_V(k))}$ for any choice of $k \geq 0$.
%% The error bound can then be defined by a min-max principle
%% \begin{align}
%% \ln(1/\epsilon) = \min_V \max_{k \geq 0} (k - \lambda_V(k)). 
%% \end{align}
%% The Taylor series of the SCGF gives the $n$-th cumulants of $N_V$, $\lambda_V(k)= |V|^{-1} \sum_{n=1}^\infty \langle N_V^n \rangle_c \frac{k^n}{n!}$. 
%% The cumulants are in turn related to the connected correlations, e.g., $\langle N_V^2 \rangle_c =\sum_{x, y \in V} \langle  E_x E_y \rangle_c$.
%% If the correlations decay, we expect the cumulants to scale as $\langle N_V^n \rangle_c = |V| c_n(V)$, where $c_n(V)$ depends on the geometry of $V$ but does not grow with $|V|$ (in particular, $c_1(V) = P_E$).
%% Let us suppose that their growth is upper bounded by a constant $c_n \equiv \max_V c_n(V)$.

\begin{figure}[t]
\includegraphics[width=3.2in]{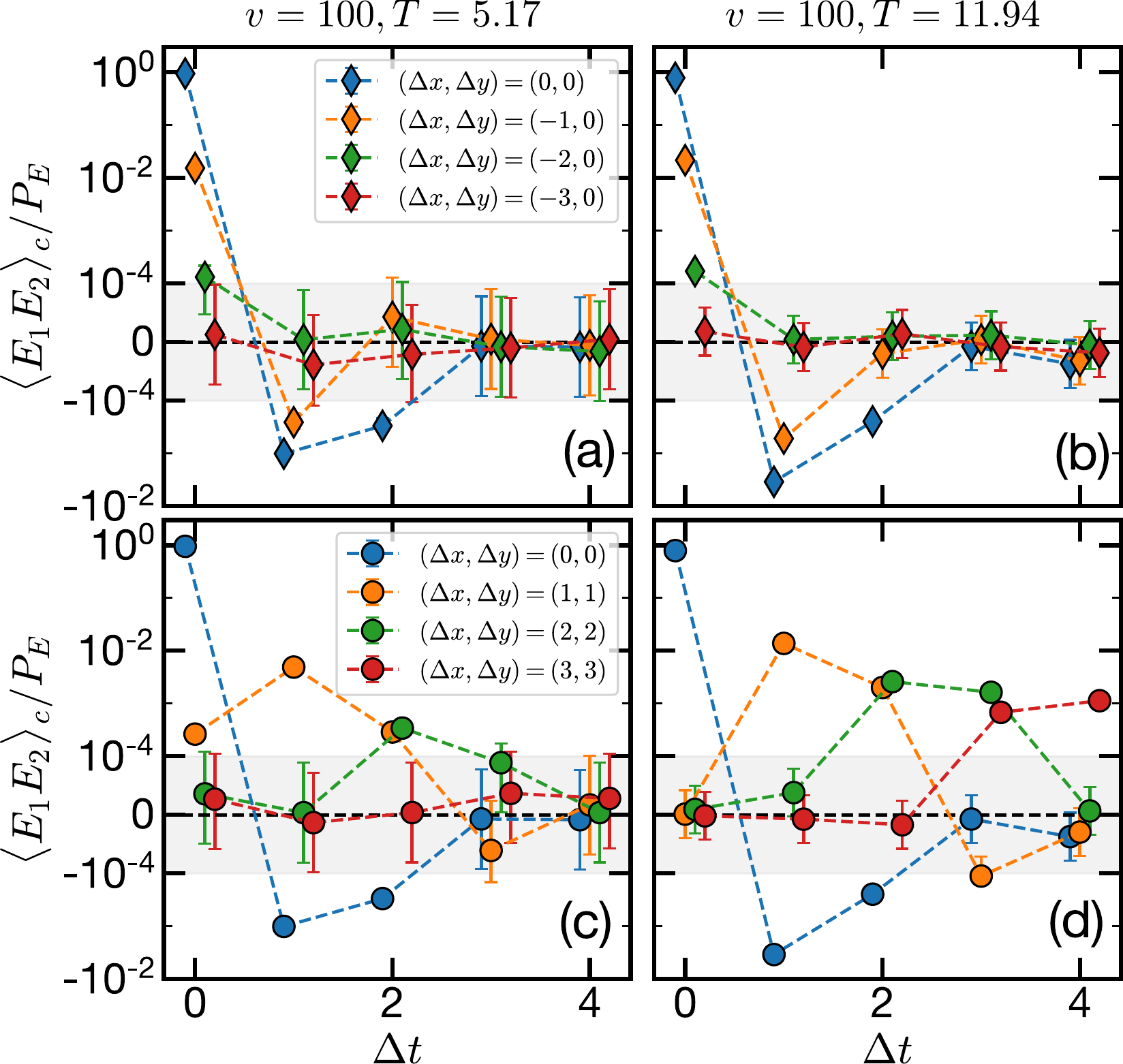}
\caption{
  Connected two-point correlations of the errors for $v=100$ and $T=5.17$ and $T=11.94$.
  Along different directions, the errors display fundamentally distinct correlations, which are more prominent for higher temperature.
}
\label{app1_SpaceTemp} 
\end{figure}

We now discuss whether the error bound $\epsilon(T) \to 0$ as $T \to 0$, ensuring that by reducing the temperature of the external bath, the error probability can be made arbitrarily small.
One sufficient condition is the existence of a $T$-independent, continuous, and strictly increasing function $\Lambda(k)$ such that $\lambda(k) \leq P_E(T) \Lambda(k)$  for all $k, T \geq 0$ and with $\Lambda(0) = 0$.
To see why, note the aforementioned min-max principle now becomes:
\begin{align}
\log\left [1/ \epsilon(T) \right] = \max_{j\geq 0} \left[j - P_E(T) \Lambda(j) \right].
\label{eq:epsT_bound}
\end{align}
Since $\Lambda(j)$ is invertible on $j\in \mathbb{R}^+$, we may define $j_\ast(P_E) = \Lambda^{-1}(1 / P_E)$. Eq.~\eqref{eq:epsT_bound} then provides the bound $\log[1/ \epsilon(T)] \geq j_\ast(P_E(T)) - 1$.
Finally, note $\lim_{P_E \to 0} j_\ast(P_E) = \infty$, because the inverse of a strictly increasing function is itself strictly increasing. 
Thus, the existence of such a $\Lambda(j)$, combined with our earlier evidence that $\lim_{T \to 0} P_E(T) = 0$, would imply $\lim_{T \to 0} \epsilon(T) = 0$.

\begin{table}[b]
  \caption{Fitting parameters for power-law fit}
  \label{Table_T11}
  \begin{tabular}{ c || c c c | c c c |}
    & \multicolumn{3}{|c|}{\makecell{$T=5.17, v=100$\\Fig.~\ref{fig:piToom_Error_Corr}}} & \multicolumn{3}{|c|}{\makecell{$T=11.94, v=100$\\Fig.~S4}}\\ \hline \hline
    $n$ & $c_n$ & $b_n$ & $\eta_n$ & $c_n$ & $b_n$ & $\eta_n$ \\ 
    \hline
    1 & 0.048 & 0 & --- & 0.21 & 0 & ---  \\
    \hline
    2 & 0.052 & 0.007 & 0.60 & 0.26 & 0.09 & 0.23  \\
    \hline
    3 & 0.067 & 0.026 & 0.46 & 0.11 & --- & ---\\ 
    \hline
    4 & 0.088 & 0.060  &  0.90  & --- & --- & ---
  \end{tabular}
\end{table}

To verify the existence of such a $\Lambda(j)$, it would be sufficient to show that the scaled cumulants are bounded as $c_n(T) \leq P_E(T) C_n $, with $C_n$ growing slower than $n!$, so that $\Lambda(j)$ has an infinite radius of convergence (Poisson statistics corresponds to $C_n = 1$, $\Lambda(k) = e^k$).
%However, due to the small statistics, numerically estimating $c_n$ at low temperatures is extremely demanding.
A preliminary comparison of $T = 5.17, 11.94$ (Fig.~3 in the main text and Fig.~4 in the supplemental materials~\cite{supp}) finds  $c_2(T) / P_E(T) = 1.24$ at $T = 11.94$, while $c_2(T) / P_E(T) = 1.08$ at $T = 5.17$, consistent with an approach to $C_2 \sim 1$, but  a comprehensive investigation remains a work in progress (summary of the extracted cumulant fits can be found in Table~\ref{Table_T11}).

We end this appendix by complementing the data presented in Fig.~3(b) of the main text with additional space-time cuts as well as different temperatures [Fig.~\ref{app1_SpaceTemp}].
With regards to the space-time cuts, we observe that the correlations in errors are clearly anisotropic---when considering cuts in the negative $\hat{x}$ direction (Fig.~\ref{app1_SpaceTemp} top row), the correlations quickly become zero once the distance between the sites is greated than 2.
On the other hand, when moving in the NEC direction, we find that error correlations are much greater and display a light-cone-like behavior---namely, errors a distance $l$ away become meaningful after $\Delta t=l$ steps has passed.
This fact is further enhanced upon increasing the temperature of the system. When going from $T\sim 5$ to $T\sim 11$, we observe that the correlated errors survive to much later times and spread to longer distances. These behaviors highlight the non-Markovianity of the noise in the Floquet-Langevin dynamics.
Nevertheless, our present numerical results suggest that the noise satisfies the condition encoded in Eq.~\eqref{error_condition}.

\bibliography{myref}

\begin{thebibliography}{77}
\bibitem [{fn7()}]{fn7}%
  \BibitemOpen
  \href@noop {} {}\bibinfo {note} {For each temperature, we examine the
  ensemble of all simulated Langevin trajectories and count the number of
  errors under the continuous to discrete mapping $S$; this allows us to
  empirically determine the error rate, $P_E(T)$.}\BibitemShut {Stop}%
\bibitem [{fn1()}]{fn10}%
  \BibitemOpen
  \href@noop {} {}\bibinfo {note} {We note that due to the spatio-temporal
  correlation of errors, the error rate is not the same as the error bound,
  $P_E \neq \epsilon$, but $P_E$ will nevertheless play an important role in
  the analysis.}\BibitemShut {Stop}%
\bibitem [{fn6()}]{fn6}%
  \BibitemOpen
  \href@noop {} {}\bibinfo {note} {Without loss of generality, we set the mass
  of the oscillator to be $m=1/2$ in the remaining discussions.}\BibitemShut
  {Stop}%
\end{thebibliography}

\end{document}

\title{Supplementary Material: Absolutely stable time crystals at finite temperature}
\author{Francisco Machado}
\thanks{These two authors contributed equally}
\affiliation{ITAMP, Harvard-Smithsonian Center for Astrophysics, Cambridge, Massachusetts, 02138, USA}
\affiliation{Department of Physics, Harvard University, MA 02138, USA}
\affiliation{Department of Physics, University of California, Berkeley, CA 94720, USA}
\affiliation{Materials Science Division, Lawrence Berkeley National Laboratory, Berkeley, CA 94720, USA}
\author{Quntao Zhuang}
\thanks{These two authors contributed equally}
\affiliation{Department of Physics, University of California, Berkeley, CA 94720, USA}
\affiliation{James C. Wyant College of Optical Sciences, University of Arizona, Tucson, Arizona 85721, USA}
\affiliation{Ming Hsieh Department of Electrical and Computer Engineering \& Department of Physics and Astronomy, University of Southern California, Los Angeles, CA 90089, USA}
%\email{zhuangquntao@gmail.com}
\author{Norman Y. Yao}
\affiliation{Department of Physics, Harvard University, MA 02138, USA}
\affiliation{Department of Physics, University of California, Berkeley, CA 94720, USA}
\affiliation{Materials Science Division, Lawrence Berkeley National Laboratory, Berkeley, CA 94720, USA}
%\email{norman.yao@berkeley.edu}
\author{Michael P. Zaletel}
\affiliation{Department of Physics, University of California, Berkeley, CA 94720, USA}
\affiliation{Materials Science Division, Lawrence Berkeley National Laboratory, Berkeley, CA 94720, USA}
%\email{mikezaletel@berkeley.edu}

%% \author{Francisco Machado$^{1,2}$}
%% \thanks{These two authors contributed equally}
%% \author{Quntao Zhuang$^{1,3,4}$}
%% \thanks{These two authors contributed equally}
%% %\email{zhuangquntao@gmail.com}
%% \author{Norman Y. Yao$^{1,2}$ }
%% %\email{norman.yao@berkeley.edu}
%% \author{Michael P. Zaletel$^{1,2}$}
%% %\email{mikezaletel@berkeley.edu}
%% \affiliation{$^1$Department of Physics, University of California, Berkeley, California 94720, USA
%% \\
%% $^2$Materials Science Division, Lawrence Berkeley National Laboratory, Berkeley, California 94720, USA
%% \\
%% $^3$Department of Electrical and Computer Engineering \&
%% James C. Wyant College of Optical Sciences, University of Arizona, Tucson, Arizona 85721, USA
%% \\ 
%% $^4$
%% Ming Hsieh Department of Electrical and Computer Engineering 
%% \& 
%% Department of Physics and Astronomy, University of Southern California, Los
%% Angeles, California 90089, USA
%% }
\date{\today}

\maketitle

\section{Time crystalline order in a 1D PCA -- the $\pi$-Gacs construction}

As with conventional phases of matter, the possibility of a stable, discrete time crystal depends on spatial dimension.
Thus, it is natural to ask if a PCA time-crystal can also exist in 1D.
Note that in the quantum case, while the 1D MBL discrete time crystal is fairly well established~\cite{khemani2016phase,else2016floquet,yao2017discrete,randall2021observation,mi2021observation}, the stability of MBL in 2D, and hence the existence of a 2D MBL DTC phase, remains controversial~\cite{de2017stability,crowley2020avalanche}.

Unfortunately, Toom's route to stability, discussed in the main text, cannot be generalized to 1D.
Since each island of errors is only separated by two domain walls in one dimension, locally one cannot efficiently tell which side corresponds to the error  and which side to the correct region.
This intuition is not specific to the Toom model, and in fact for many decades it was conjectured that all 1D, finite-range and finite-state PCAs were generically ergodic, i.e.~the so-called ``positive-rates conjecture''~\cite{gray2001reader}.

Surprisingly, in 1998, this longstanding conjecture was proven incorrect by G{\'a}cs~\cite{gacs2001reliable}.
G{\'a}cs constructed a 1D translation invariant PCA, with nearest neighbor interactions, with the following remarkable property: on a chain of length $L$, the dynamics exhibit $2^L$ stable stationary measures (intuitively, one can think of these as fixed points) in the limit $L \to \infty$; said another way, G{\'a}cs' PCA can ``remember'' one bit per unit length! 
Each cell/site of the PCA has a large state space, likely somewhere between $2^{24}$ and $2^{400}$~\cite{gray2001reader,gacs2001reliable}. Roughly speaking,  each cell contains one bit  that it is trying to remember, and the remaining 399 bits (taking e.g.~the $2^{400}$ state space) are involved in a highly collective error correction protocol. 
As in the Toom model, the stochastic errors can be  biased so long as they remain below some finite threshold,  above which  a dynamical phase transition will restore ergodicity.

Even more remarkably, not only is the G{\'a}cs model an error-corrected memory, it can execute Turing-complete operations on the protected state space. In other words, his construction demonstrates that a \emph{stochastic} 1D PCA can be used to simulate a deterministic CA, and hence error-corrected classical computing is possible in 1D. 
This immediately implies the existence of the ``$\pi$-G{\'a}cs time crystal:'' In particular, one can simply use the G{\'a}cs construction to emulate a CA with the rule: $1 \leftrightarrow -1$. His mathematical results then  imply that this is an absolutely stable discrete time crystal, with infinitely long-lived temporal order as  $L \to \infty$.

G{\'a}cs' result  is mathematically rigorous, and as such, there are assumptions about the error model~\cite{toom1980stable, gacs2001reliable, gacs2021new}.
Crucially, as discussed in the main text, this construction is stable to very generic noise models: the noise need not be Markov, and needs only to satisfy:
\begin{align}
P_{\wedge_{\ell=1}^k E_{u_\ell}}=\prod_{\ell=1}^k P_{E_\ell|E_{\ell-1}\cdots E_1}\le \epsilon^k,
\label{error_condition}
\end{align}
where $E_{u_i}$ denotes the indicator function of whether an error has occurred at the space-time point $u_\ell$.
Thus, a sufficient condition is simply to require that each $P_{E_\ell|E_{\ell-1}\cdots E_1}\le \epsilon$; if we choose $\ell$ to be ordered in time, this will be satisfied if the probability for an error to occur at $\ell$, conditioned on all of the past errors, is below some constant $\epsilon$.
This will \emph{always} be the case for a  Markov model of the form, $M = \mathcal{T} + \epsilon \Delta M$, for arbitrary (local) entries  $|\Delta M_{\eta \to \eta'}| < 1$. Thus, the ``stability to errors'' of the G{\'a}cs model can be understood, more generally, as the stability to  local perturbations.

As a result, as long as the error can be made small enough and satisfy Eq.~\eqref{error_condition}, one has a path towards building an absolutely stable one dimensional time crystal.

\section{Details of the Floquet Langevin Evolution and Numerical Simulation}

\subsection{More detailed description of the mapping} 

\begin{figure}[t]
  \includegraphics[width =0.8\textwidth]{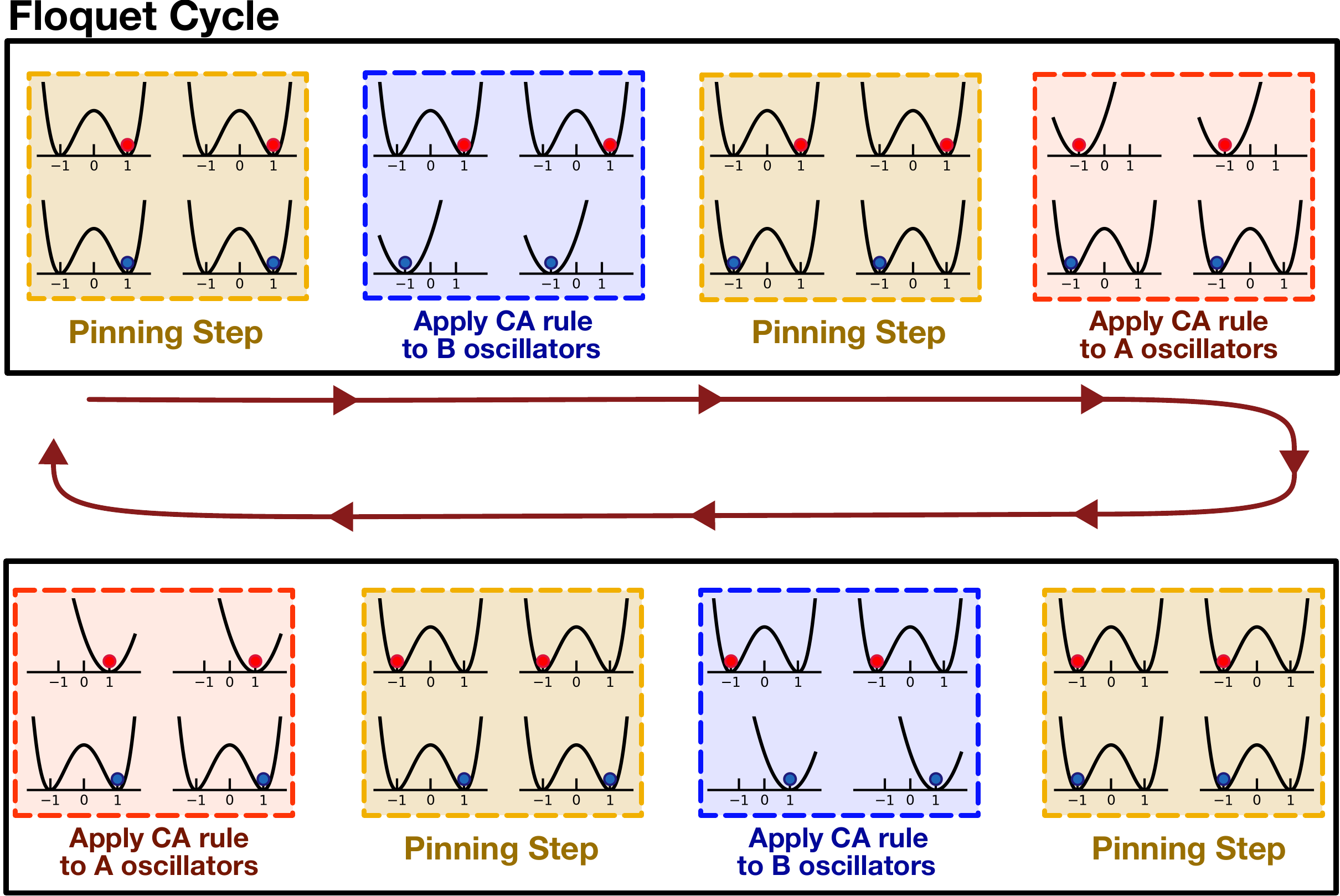}
  \caption{Schematic of two $\pi$-Toom Floquet cycles using our protocol. In steps 1 and 3, a pinning potential is applied to allow the bath to extract energy from the system. In step 2, the CA rule is applied on the oscillators $B$ based upon the state in oscillators A; in step 4, the role of the two types of oscillators is reversed.} 
  \label{fig:ExpSteps}
\end{figure}

In this section we discuss in some additional detail our construction for mapping the CA dynamics within a noisy Floquet-Langevin classical system.
We begin with the oscillators at site $\bm{x}$ in the state $(q^{A/B}_{\bm x}, p^{A/B}_{\bm x}) = \Big(Q\big(\eta(\bm{x},t)\big), 0\Big)$, where $Q(s)$ is the function that maps a particular CA state to a position of the oscillator. Given the discrete nature of the $\mathcal{S}$, one can map each state $s\in \mathcal{S}$ to a different integer in the real line of positions.
From there, the dynamics evolve via a 4-step process, Fig.\ref{fig:ExpSteps}. 

\begin{figure}[t]
\centering
\includegraphics[width=0.495\textwidth]{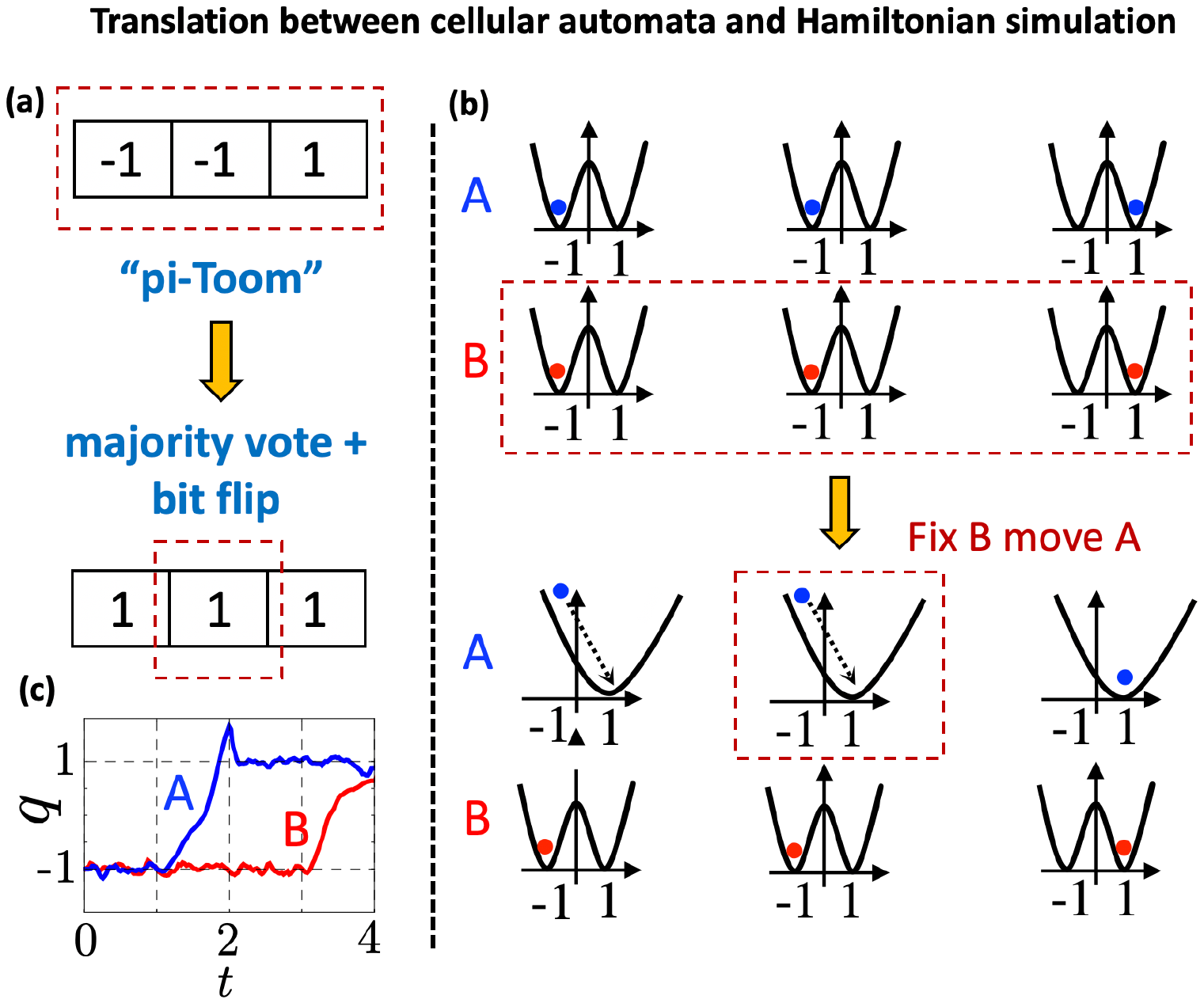}
\caption{Schematic depicting the translation between a cellular automata step and the corresponding Hamiltonian simulation step.
(a) Shows a single step of a one dimensional version of the ``$\pi$-Toom'' rule, which consists of a majority vote and a bit flip. 
(b) In the Hamiltonian setting, we consider two sets of oscillators, A (blue) and B (red).
The corresponding Hamiltonian simulation proceeds in two steps. First, there is a ``relaxation'' step followed by local interactions which implement the ``majority vote''. In this second step, the B oscillators are fixed, while the state of the A oscillators is updated.
 (c) Trajectory $q_{A/B}(t)$ during an error correction step of  the Toom model for a $32\times 32$ lattice with $v=50$ and $T=2$.
  The vertical dashed lines divide time into four steps as described in the text;  an error of the form $q\sim-1$ gets corrected to be $q\sim 1$. 
\label{Hmodel_schematic}
}
\end{figure}

\vspace{2mm}

\textbf{Step 1: Relaxation.} The goal of the first step is to leverage  dissipation  in order to reduce fluctuations in the system.
In particular, we envision turning on a one-body potential, $V_{\textrm{pin}}(q)$, which has a local minimum at $Q(s)$ for all $s\in\calS$.
At sufficiently low temperatures, the dissipative dynamics will relax the oscillator's positions, $q_{\bm x}$, toward valid values of $Q(s)$ with low momenta (with fluctuations of order  the equipartition scale $\sim k_B T$). 
The precise form of $V_{\textrm{pin}}$ is not important; however, for concreteness we will utilize
\begin{align}
V_{\textrm{pin}}(q) =  v_{\textrm{pin}} \prod_{s \in \mathcal{S}} (Q(s) - q)^2  
\end{align}
where the overall magnitude of the pinning potential is set by $v_{\textrm{pin}}$.

\vspace{2mm}

\textbf{Step 2: Fix A move B.} As illustrated in Fig.~\ref{Hmodel_schematic}, the second step of the Floquet dynamics implements the cellular automata transition $B = \mathcal{T}(A)$.
We will keep $q^A$ fixed using the pinning potential, $V_{\textrm{pin}}$. 
For the $B$ oscillators, however, we turn off $V_{\textrm{pin}}$, and turn on an interaction $V_I$ between $q^A$ and $q^B$.
This interaction is engineered such that each $q_{\bm x}^B$ sees only a single potential minimum corresponding to the desired CA update rule; in general, this will depend on the state of the $A$ oscillators in the associated neighborhood, $\{q_{\bm{x}+\calN}^A\}$. 

Defining the location of this minimum to be $\widetilde{\calT}(\{q_{\bm{x}+\calN}^A\})$, we can then specify an interaction of the form:
\begin{align}
  V_{I}(\{q^A_{\bm{x}+\calN}\}, q^B_{\bm x}) = \frac{v_I}{2} \left( \widetilde{\calT}(\{q^A_{\bm{x}+\calN}\}) - q^B_{\bm x}\right)^2,
  \label{VI_implementation}
\end{align}
where the interaction strength is characterized by  $v_I$.
This is a highly non-linear but local interaction between each $q^B_{\bm x}$ and a finite set of $A$ oscillators, $\{ q^A_{ {\bm x} + \mathcal{N}} \}$, within the neighborhood, $\mathcal{N}$.
In particular, as shown in Fig.~\ref{Hmodel_schematic}(b), for the example of $\widetilde{\calT}$ being an ``anti-majority vote'', the interaction would correspond to an  $|\mathcal{N}|+1$ body coupling~\footnote{Naively, one could set $\tilde{\mathcal{T}}(\{q_{\bm{x}+\calN}^A\}) = Q\Big(\calT\big( S(q_{\bm{x} + \calN}^A)\big)\Big)$.
However, the discontinuities in $S(q)$, and hence $V_I$, lead to isolated points with infinite force, which makes any analysis or numerical simulation significantly more troublesome.
To remedy this, we smooth the interaction using an interpolation as detailed in the section \emph{Smoothing discontinuities}.}.

\vspace{2mm}

\textbf{Step 3: Relaxation.} In the third step,  we turn off the interaction, $V_I$, while ramping up the pinning potential, $V_{\textrm{pin}}$.
As in the first step, dissipation relaxes and pins the positions of the oscillators.

\vspace{2mm}

\textbf{Step 4: Fix B move A.} In the final step, we implement ``$A = \mathcal{T}(B)$'' by repeating step two with the role of $A$ and $B$ reversed. 

\vspace{2mm}

After these four steps, our Floquet dynamics  have implemented two steps of the cellular automata update rule, $\cal T$.
This block naturally forms a single period of the Floquet drive, which can then  be repeated.
As mentioned in the main text, one can also replace the transition $\mathcal{T}$ in the last step with the ``do-nothing'' CA rule  if one wants to implement only a single CA update, $\mathcal{T}$, per Floquet cycle.

\subsection{Estimating temperature dependence of error $P_E$}
To predict the rate of such errors, suppose that the interaction potential is driving an oscillator to the state $q=-1$, so that $V_I(q)=\frac{v_I}{2} (q+1)^2$. 
When the dynamics switch to the pinning potential $V_{\textrm{pin}}(q)$, an error will occur if the oscillator has a position  $q \in [0,\infty)$.
Assuming the system reaches local equilibrium with respect to $V_I$, the probability of this error can be estimated from the Boltzmann distribution as:
\begin{align}
P_E\left(\frac{v_I}{T} \right) \equiv \frac{\int_0^\infty e^{-V_I(q)/T}}{\int_{-\infty}^\infty e^{-V_I(q)/T}}=\frac{1}{2}{\rm Erfc}\left (\sqrt{\frac{v_I}{2T}} \right),
\label{pe_equi}
\end{align}
which asymptotically gives exponential decay $P_E \sim e^{-v_I / 2 T}$.
This is the functional form used in Fig.~2(a) of the main text, in excelent agreement with the numerical data.

\subsection{Effect of the choice of damping parameter $\gamma$}

\begin{figure}[t]
\includegraphics[width=0.5\textwidth]{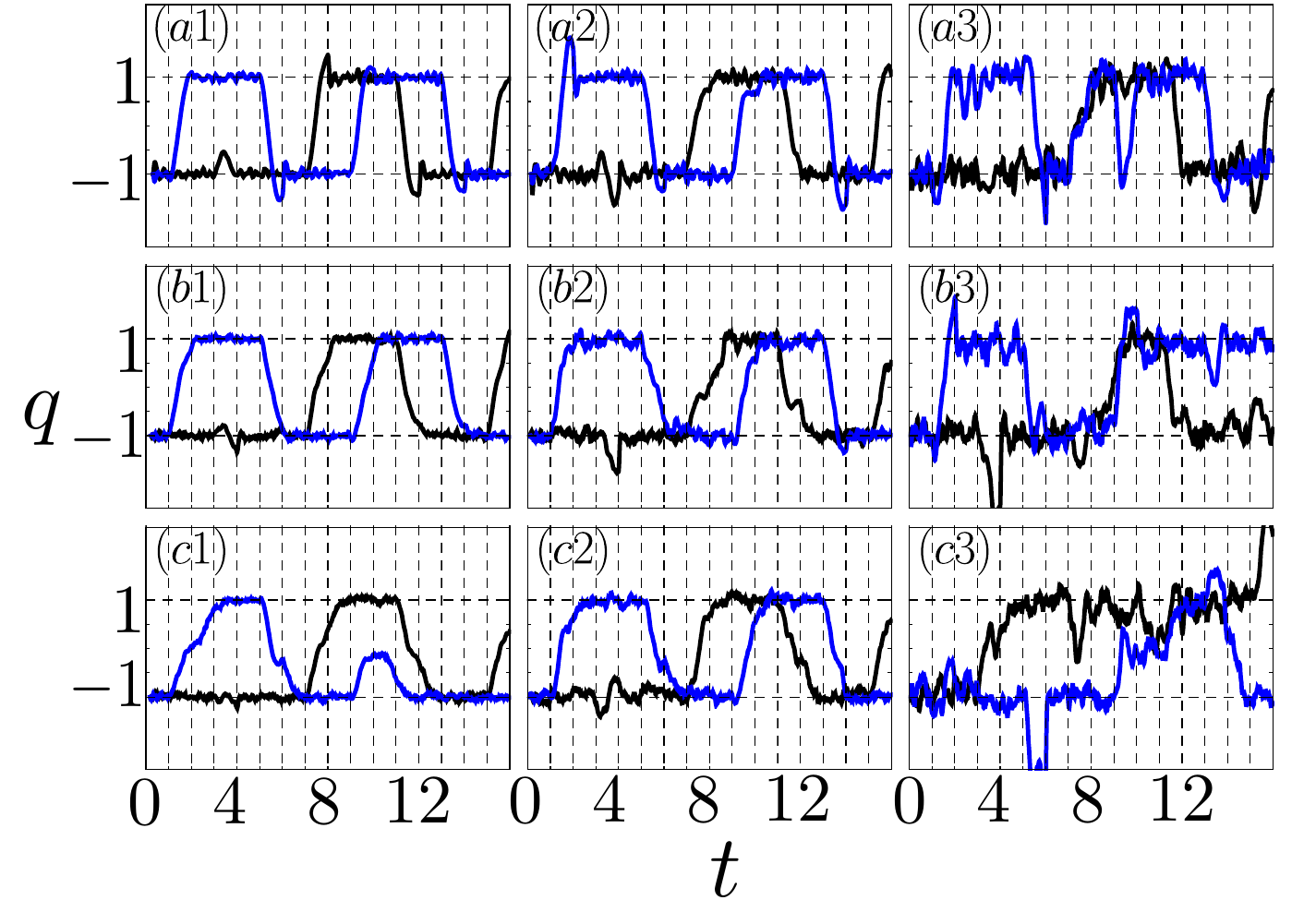}
\caption{
Error correction process in Floquet Langevin simulation of the $\pi-$Toom model. Dynamics are shown for a $32\times 32$ lattice with a pinning potential $v=50$. The blue lines represent the $A$ oscillator and the black lines represent the $B$ oscillator. From left to right, the temperature is varied such that $T=0.5,2,10$. From top to bottom, we change the damping ratio, $\kappa_f=0.5,1,1.5$
\label{app1_piToom}
}
\end{figure}

In the ``fix $A$, move $B$'' step, if we fix the $A$ oscillators, then the $B$ oscillators are critically damped harmonic oscillators with additional noise. 
If there is no noise, then replacing $q_k^B=\exp\left(\lambda t\right)$ gives two eigenvalues $\lambda=-(1/2)\left(\gamma \pm \sqrt{\gamma^2-4v_I}\right)$.
Thus, the decay of oscillation amplitude is fastest when the damping has the critical value
$
\gamma^\star=2\sqrt{v_I}.
$
When $\gamma>\gamma^\star$, it is over-damped, and the relaxation is slow because the velocity is low. When $\gamma<\gamma^\star$,  oscillation remain and the damping is slow due to small friction coefficient.
Similarly, in the ``relaxation'' steps, both oscillators correspond to damped harmonic oscillators for the potential $V_{\textrm{pin}}(q)=v_{\textrm{pin}} \left(q^2-1\right)^2\simeq 4v_{\textrm{pin}} (q-1)^2$ near $q=+1$. The critical damping during the relaxation step is then given by $\gamma_r^\star=4\sqrt{2v_{\textrm{pin}}}$. We tune the damping ratio $\kappa_f=\gamma/\gamma^\star=\gamma_r/\gamma_r^\star$ such all steps within the Floquet evolution are tuned to critical damping. 
For simplicity, we let $v_{\textrm{pin}}=4v_I=v$, and then the critical relaxation time when $\kappa_f=1$ is $t_{r}^\star\sim 1/\sqrt{v}$. The critical damped solution when the initial oscillator position is $q_0$ at rest is given by $q(t)=q_0e^{-\sqrt{v_I}t}(\sqrt{v_I} t+1)$.

We now study whether this construction can effectively correct implement CA rules. To this end, we study the correction of a single error under $\pi$-Toom dynamics.
We set  the initial position of the $A$, $B$ oscillators to be uniformly $q=+1$ in all cells, except at cell $(1,1)$, where we choose $q = -1$ to create an error.
All initial momenta are zero.
In Fig.~\ref{app1_piToom}, we monitor the position  of the $A$ and $B$ oscillator at  cell $(1,1)$. 
In this plot, between $[0,1],[2,3],[4,5],\cdots$ we have the relaxation steps; between $[1,2],[5,6],[9,10], \cdots$, we have Toom steps; while between $[3,4],[7,8],[11,12],\cdots$ we have $\pi$-Toom steps.
We see that in all cases the error gets corrected within a duration of $t=4$, after which the black and blue curves have similar shapes at low temperature; however, the critical damping case [Fig.~\ref{app1_piToom}(b1-b3)] relaxes the fastest and exhibits the smallest fluctuations, compared with the under-damped [Fig.~\ref{app1_piToom}(a1-a3)] and over-damped [Fig.~\ref{app1_piToom}(c1-c3)] cases.
We have verified that the same behavior holds for various pinning potentials $v$ and  temperatures $T$.

\subsection{Smoothing discontinuities}
\label{sec:smooth}
The interaction $V_I$ in Eq.~\eqref{VI_implementation} involves two set of coordinates: $q^B_{\bm x}$ and $\{q^A_{\bm{x}+\calN}\}$. In the numerical simulation, one needs to implement the motion by applying a force on the $A$ oscillators and $B$ oscillators. 
Naively, the $\widetilde{\mathcal{T}}$ of Eq.~\eqref{VI_implementation} would be taken as $Q(\mathcal{T}(S(q)))$.
However, due to the discontinuities in $S$, this choice would cause $V_I$ to depend discontinuously on $\{q^A_{\bm{x}+\calN}\}$, leading to $\delta$-function forces.
This would make the numerical simulation of Hamilton's equations challenging.
To avoid this, we first evaluate the values of $V_I$ at a discrete set of points of $q^A_{\bm{x}+\calN}=\pm1$, while keeping $q^B_{\bm x}$ arbitrary. We then create a linear interpolation between these discrete points to create a smoothed version of $V_I$ for oscillators $A$, which allows the evaluation of the force in the numerical simulation.

\section{Effect of initial state on the Floquet-Langevin $\pi$-Toom dynamics}
\begin{figure}
\centering
\includegraphics[width=0.85\textwidth]{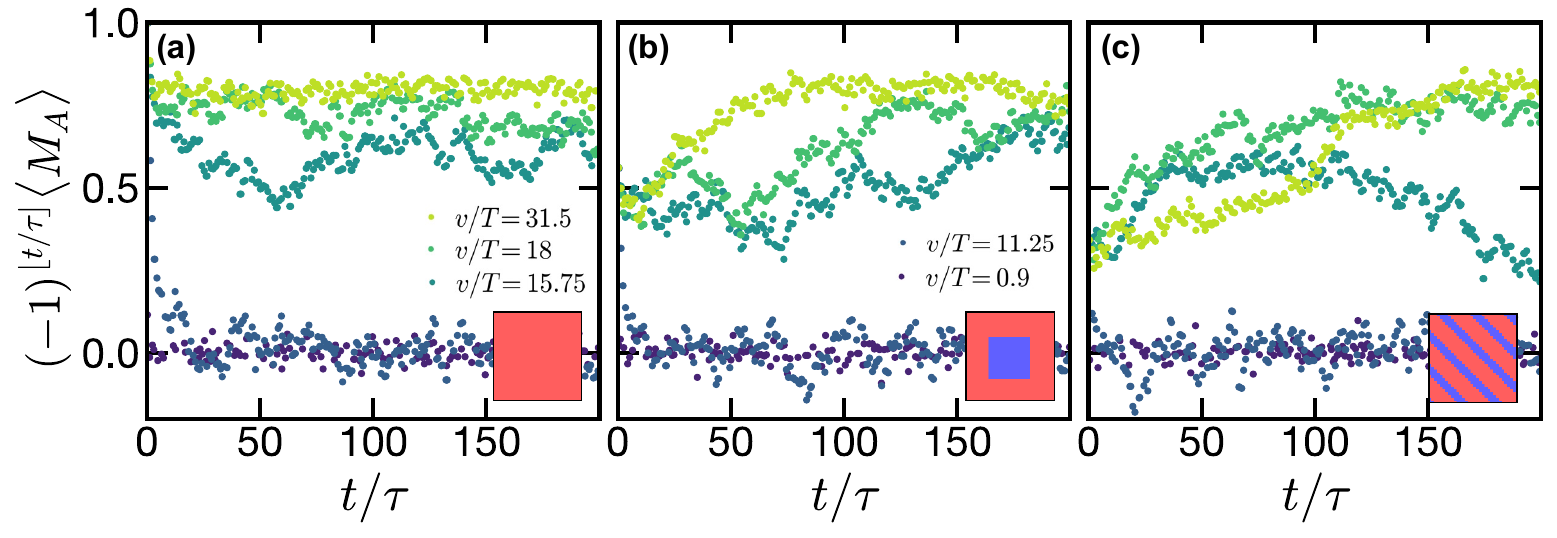}
\caption{Floquet Langevin simulation of the $\pi$-Toom model for a two dimensional lattice of size $32\times 32$, with $v=50$. For each panel, the inset depicts the initial configuration (for both $A$ and $B$ oscillators), where red indicates $q=+1$ and blue indicates $q=-1$.
(a) For a uniform initial state, at low temperatures, the time crystalline order parameter remains finite at late times. At high temperatures, the time crystal quickly melts into a disordered phase. (b) For an initial state with a central island of errors (i.e.~oscillators in $q=-1$), at low temperatures, the Floquet Langevin dynamics ``error correct'' and the time crystalline order grows toward a plateau at late times. At high temperatures, the time crystal again melts into a disordered phase. (c) For an initial state with stripes of errors, one sees the same qualitative behavior as in panel (b). As expected, near the transition (data set with $v/T=15.75$), it becomes difficult to tell whether the time crystalline order will eventually decay or plateau to a finite value.
\label{dtc_toom_dynamics}
}
\end{figure}

To investigate the emergence of DTC order, we compute the Floquet-Langevin dynamics starting from three distinct initial states: (i) a uniform  input state with all oscillators in $q=+1$ [Fig.~\ref{dtc_toom_dynamics}(a)], (ii) a state which contains an island of $q=-1$ oscillators in the center [Fig.~\ref{dtc_toom_dynamics}(b)], and (iii) a state which consists of diagonal stripes of $q=-1$ oscillators [Fig.~\ref{dtc_toom_dynamics}(c)] \footnote{Note that for perfect stripes, even the deterministic Toom model cannot not correct it. However, there are a measure zero set of such fine-tuned initial conditions.}.
 In the language of the $\pi$-Toom PCA, for each of these initial states, one can think of the oscillators  with $q=-1$ as ``errors'', which will either be ``corrected'' by our Floquet Langevin dynamics (for sufficiently low bath temperatures) or not.

For the uniform initial state [Fig.~\ref{dtc_toom_dynamics}(a)], the DTC order parameter, $(-1)^{\lfloor t/\tau \rfloor} \braket{M_A}$, begins at unity for all temperatures. At high temperatures, the order parameter quickly decays to zero, indicating that the Floquet Langevin dynamics drive the system toward the disordered phase. 
On the other hand, for sufficiently low temperatures, the time crystalline order evolves toward a finite plateau value at late-times, indicative of a DTC. 
In Fig.~\ref{dtc_toom_dynamics}(b), we show the analogous dynamics starting from an initial state with a central island of errors. 
For low temperatures, the Floquet Langevin dynamics correct these errors and the DTC order parameter grows, with the system approaching a  time crystalline state.
Again, above a critical temperature, time crystalline order ``melts'' and the stroboscopic magnetization decays to zero. 
Finally, Fig.~\ref{dtc_toom_dynamics}(c) depicts the dynamics starting from a striped error configuration; the qualitative features are identical to Fig.~\ref{dtc_toom_dynamics}(b), although the competition between the DTC phase and the disordered phase is more apparent at intermediate temperatures. 

\begin{figure}[t]
\includegraphics[width=0.4\textwidth]{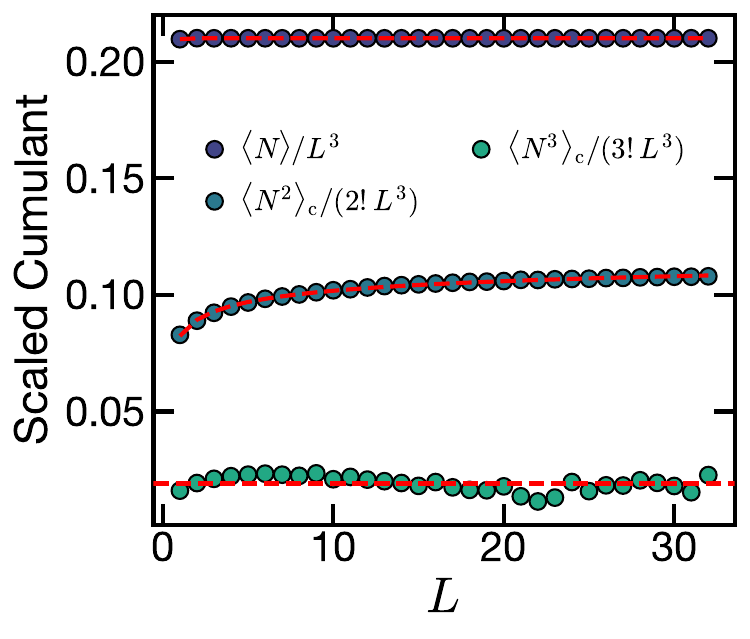}
\caption{
  Cumulants,  $\langle N^n_V \rangle_c$, of the error distribution from a Floquet Langevin simulation of the $\pi$-Toom model for $T=11.94, v=100$.
  We find that accurately estimating higher cumulants, e.g.~$c_3$ and $c_4$, is more challenging at temperatures above the critical temperature, $T_c = 9.6$.  
}
\label{app1_cumulants_T11} 
\end{figure}

\section{Finite-Size Scaling behavior of the $\pi$-Toom time crystalline dynamics}

In this section, we complement our analysis of the robustness of the $\pi$-Toom time crystal by demonstrating that the lifetime of the time crystalline dynamics scales \emph{exponentially} in system size, and does not appear to host any saturation--this further strengths our claims of the stability of the $\pi$-Toom time crystal in the $L\to \infty$ thermodynamic limit.

\begin{figure}
    \centering
    \includegraphics[width = 0.9\textwidth]{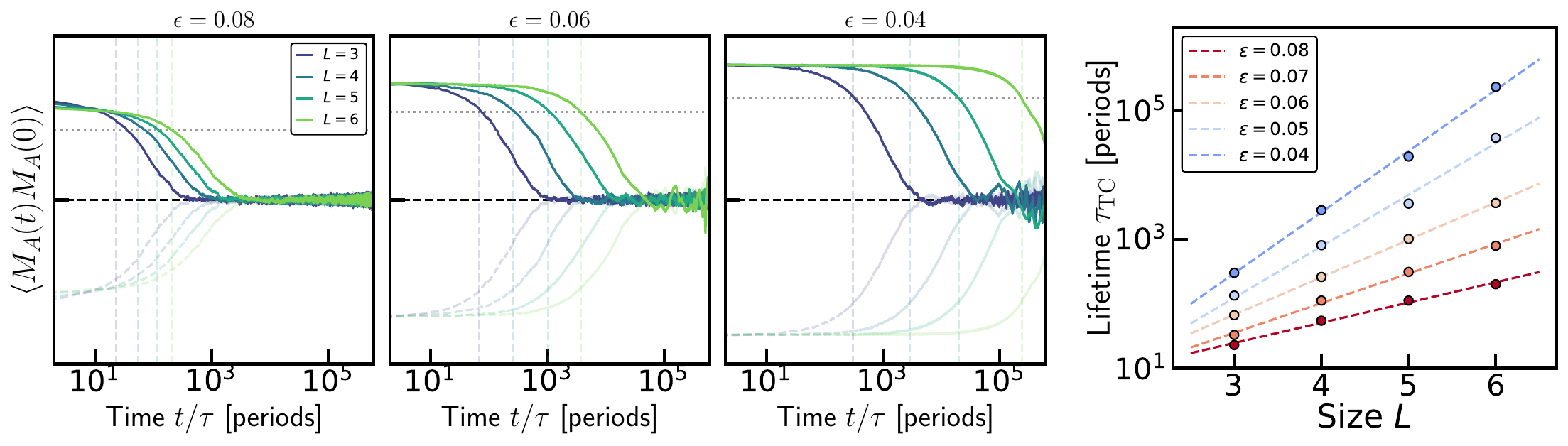}
    \caption{
    Left Panels: Dynamics of the two-point correlation function $\langle M_A(\tau)M_A(0)\rangle$ of the magnetization $M_A$ under the PCA dynamics for even (full lines) and odd periods (light dashed lines) and varying system size.
    Regardless of the choice of error rate $\epsilon$, the system exhibits a robust period doubling behavior whose lifetime \emph{grows} with the system size of the system.
    Right Panel: Dependence of the time crystalline lifetime $\tau_{\text{TC}}$ on the system size for different error rates $\epsilon$. 
    For all $\epsilon < 0.1$ within the $\pi$-Toom time crystal phase, $\tau_{\text{TC}}$ grows \emph{exponentially} with system size, and does not exhibit any saturation behavior. 
    This supports the stability of the $\pi$-Toom time crystal in the thermodynamic limit.
    }
    \label{fig:PCA_system}
\end{figure}

We begin by considering the PCA dynamics of the $\pi$-Toom model, Fig.~\ref{fig:PCA_system}.
In this case, the stability of the $\pi$-Toom model immediately inherits from the proven stability of the underlying Toom model~\cite{toom1980stable}. 
Nevertheless, its numerical analysis provides a controlled setting where we can better understand the signatures of the stability of the $\pi$-Toom time crystal when considering the full Langevin dynamics.

For different error rates $\epsilon$, we consider the PCA dynamics for different system sizes $N=L\times L$.
Crucially, we observe that by increasing the system size of the system, the lifetime of the time crystalline behavior grows as well.
This growth is also enhanced upon decreasing the error rate $\epsilon$ of the dynamics since $\epsilon$ reduces the probability of an extensive error occurring and thus destroying the synchronized behavior that is stabilized by Toom's rule.

For a more quantitative analysis of this decay, we extract the lifetime of the time crystalline behavior $\tau_{\text{TC}}$ as the time at which $\langle M_A(t)M_A(0)\rangle$ first becomes smaller than $0.75 \langle M_A(0)^2\rangle$.
Plotting this lifetime as a function of system size, we observe that the enhancement of the lifetime is exponential in the system size and does not display any signs of saturating---this demonstrates the robustness of the $\pi$-Toom time crystal within the PCA model.

\begin{figure}[t]
    \centering
    \includegraphics[width = 0.9\textwidth]{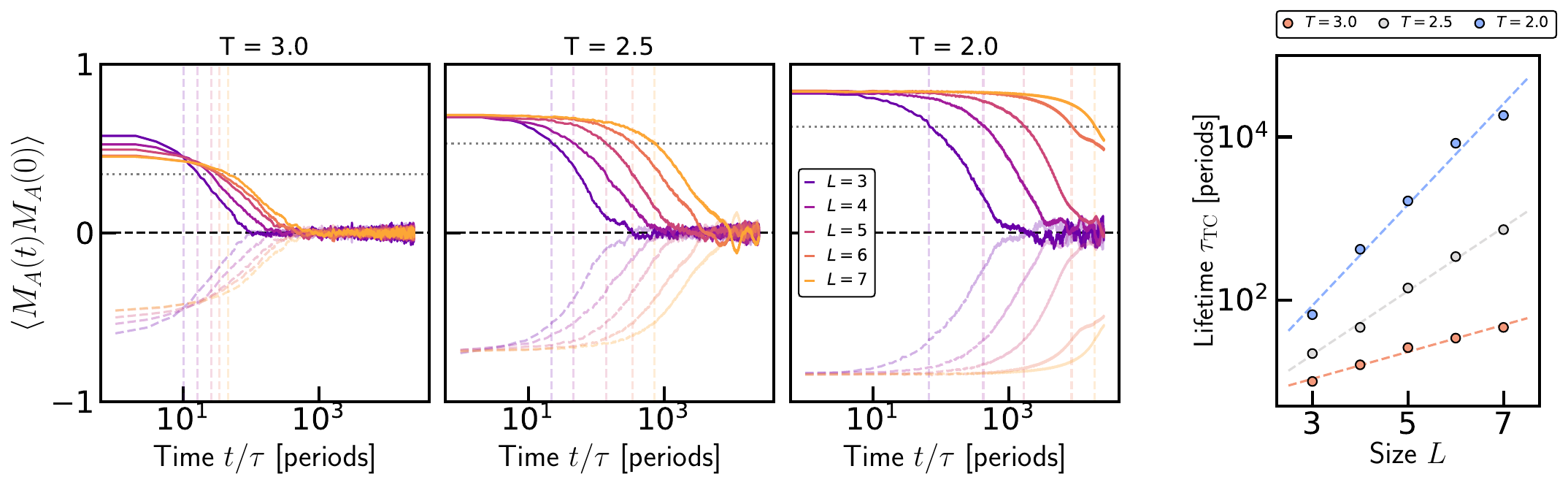}
    \caption{Analogous data as in Fig.~\ref{fig:PCA_system}, but considering the \emph{full Langevin dynamics} with $v=50$ and varying the temperature $T$ of the dynamics.}
    \label{fig:LangevinV50_system}
    \centering
    \includegraphics[width = 0.9\textwidth]{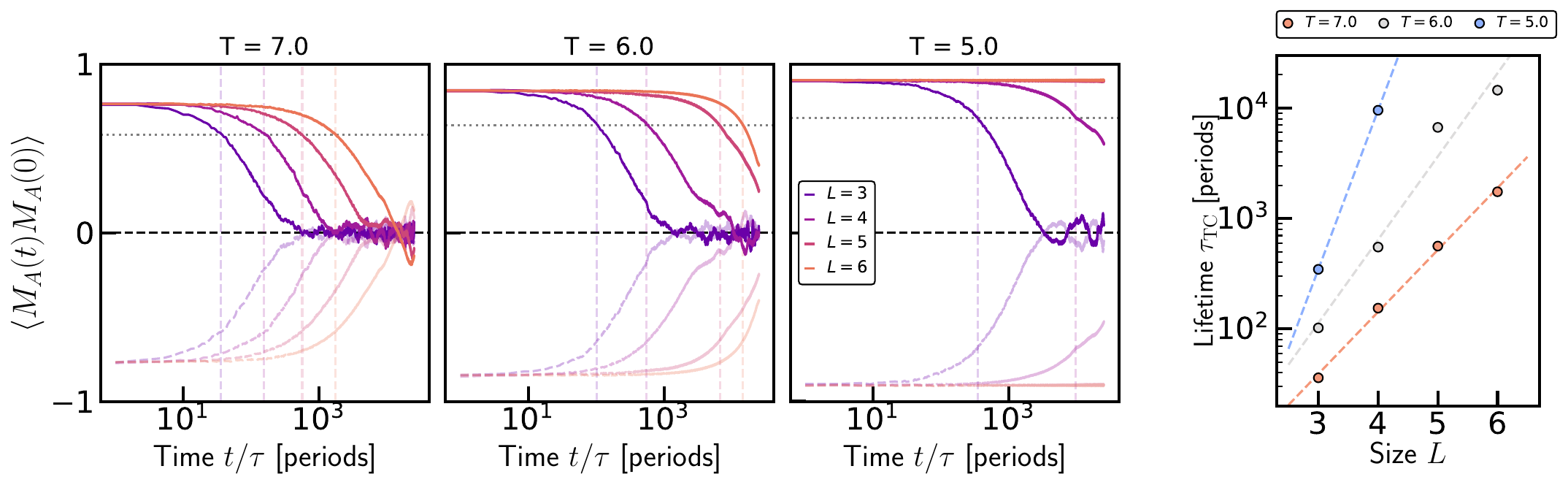}
    \caption{Analogous data as in Fig.~\ref{fig:PCA_system}, but considering the \emph{full Langevin dynamics} with $v=100$ and varying the temperature $T$ of the dynamics.}
    \label{fig:LangevinV100_system}
\end{figure}

We then perform the same analysis in the full Langevin dynamics. Owing to computational cost, we are limited to a smaller number of parameters, however we observe the same behavior, regardless of the depth of the potentials $v$, Figs.~\ref{fig:LangevinV50_system} and \ref{fig:LangevinV100_system}.
For different temperatures (at fixed potential depth), we observe that increasing the system size substantially increases the lifetime of the time-crystalline behavior and preserved the period doubling dynamics.
Crucially, upon extracting the lifetime of the time crystalline order, we observe the same exponential scaling with system size, demonstrating the robustness of the $\pi$-Toom time crystal.
Since there are no signatures of saturation, this behavior is expected to extend to the thermodynamic limit, justifying our description of the $\pi$-Toom time crystal as an out-of-equilibrium phase of matter.

% In order to show that this is not the case, we consider the \emph{system size} dependence of the time crystalline lifetime. 
% To understand why the lifetime allows us to better understand the stability of the $\pi$-Toom time crystal, it is important to note that if there are no errors in the dynamics, 

% f the the system is in the ordered phase, this error process is the only way to destroy the order and thus it sets the lifetime of the order parameter of the system.

% For any finite system size, the lifetime of the $\pi$=Toom time crystal must be finite, as there is a small, but finite error, that a majority of the sites undergo an error and the time crystalline order is destroyed. 
% However, the probability of this error becomes vanishingly small as one increases the system size; more specifically, assuming each site has an independent probability of undergoing an error, then it should be exponentially vanishing in the system size of the system $\epsilon_{err} e^{N\epsilon_{\text{eff}}}$.

% To check for this possibility we 

\newpage
\section{Experimental Realizations}

In this section, we describe two experimental routes for implementing the PCA simulation protocol described in the main text.
Before detailing these two avenues, let us recall the presented protocol, using that as a stepping stone to formalize the requirements of a suitable experimental platform~Fig~\ref{fig:ExpSteps}.
The Floquet Langevin dynamics are divided into 4 steps:
in steps (1) and (3), the potential is in a ``pinning'' configuration, whereby a strong double-well potential (in the case of a two-state cellular automata) is used to relax the oscillators and extract energy;
while during steps (2)[(4)] a multi-particle interaction is used to implement the cellular automata rule on the B[A] oscillators, based upon the state of the A[B] oscillators.
To this end, we require an experimental platform to:
\begin{itemize}
\item[{\bf (i)}] apply a non-harmonic potential to the oscillator;
\item[{\bf (ii)}] apply a multi-particle potential to each particle and its neighborhood;
\item[{\bf (iii)}] couple the system to a thermal bath.
\end{itemize}

\begin{figure}[t]
  \includegraphics[width=\textwidth]{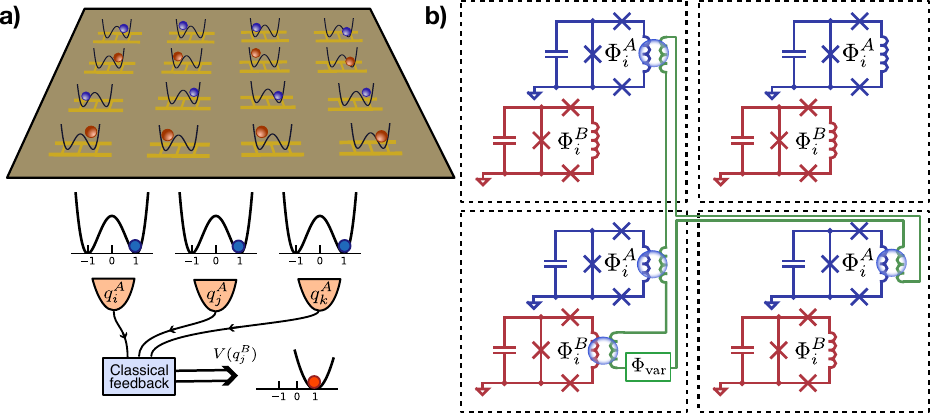}
  \caption{
    {\bf a)} An array of trapped ions serves as oscillators for the simulation of the PCA dynamics (top).
    Local electrodes can control the motional potential and dynamically change it between a pinning potential or a single-well potential centered around the positions associated with each state of the cellular automata.
    The potential necessary for implementing the PCA transition rule can be obtained by measuring the memory ions, and feed-forwarding the transition rule onto the potential experienced by the active ions (bottom).
    {\bf b)}
    An array of superconducting circuits provides a set of tunable, anharmonic oscillators, that serves as a platform for the studying the Langevin dynamics explored in the main text.
    For each cell of the PCA, one requires two circuits (blue and red), which can be coupled to each other and their neighbors (green).
    These additional coupling circuits enable the dynamically, tunable interactions necessary for implementing the PCA transition rules.
  }
  \label{fig:expt}
\end{figure}

We propose two different experimental settings where we envision our protocol being realized: ions trapped in an array of surface-electrode traps [Fig.~\ref{fig:expt}(a)]; and an array of coupled superconducting circuits [Fig.~\ref{fig:expt}(b)].

\vspace{5mm}
{\bf Ions trapped in Pauli traps}\\

In the first setting, recent developments in surface traps have enabled the exquisite control over the motional potential of ions in space~\cite{cirac:1995,kim:2010, maunz:2016,Rossnagel2016, li:2017, jain:2020, moses:2023}, offering the ability to prepare and dynamically control complicated anharmonic potentials for each individual ion~\cite{tanaka:2021a}---this meets requirement {\bf (i)}.
While electrostatic forces can be leveraged to induce pairwise interactions between the ions, such interactions would be highly dependent on the geometry of the array and, owing to their long-range Coulomb nature, not constrained to a particular finite neighborhood. As a result, ensuring that they are compatible with the cellular automata of interest requires additional considerations with respect to the geometry of the ion traps.
We propose an alternative approach, using measurements and feed-forward to dynamically generate the interaction potential $V_I$.
We envision the position of the ions being constantly monitored, and based upon the position of the ``memory'' oscillators, the potential of the ``active'' oscillators is controlled, effectively generating a multi-particle interaction potential---this meets requirement {\bf (ii)}.
We note that this approach is possible owing to the ability to dynamically control the motional potential in real-time.
Crucially, this approach is agnostic to the PCA being simulated, and thus simulating different PCAs (even embedded in different geometries and dimensions) does not require modifying the underlying hardware.
Intriguingly, by adding a time delay between the observation and the update of the potential one can also alter the nature of the system's dynamics---in this case, the system's dynamics can be made to depend explicitly on their history and not just the current position of the particles leading to non-Markovian dynamics.

Interestingly, we note that, recent progress in the engineering of multi-body interactions between the internal degrees of freedom of trapped ions hints at the possibility of implementing the required potentials through the system's native physical interactions \cite{katz:2022, katz:2023}. However, the extension of these ideas to interactions between \emph{motional} modes, as well as their incorporation with the double-well potential requires additional exploration.

Finally, we discuss requirement {\bf (iii)}. The coupling of the ions to the outside world comes through different mechanisms (i.e. black-body radiation, electric current fluctuations in the electrodes, background gas collisions), all of which can be tuned (via temperature, pressure of the chamber and further engineering of the surface electrodes)~\cite{Brownnutt2015}.
At the same time, additional noise and dissipation can be explicitly added to the system, either by adding noise to the contact potentials and thus adding fluctuations to the ion's potentials, or by using laser side-band cooling to remove kinetic energy from the ion's motion.
This level of control allows one to go beyond thermal baths to explore the effect of different spectral features of the noise on the system's ability to simulate a PCA~\cite{CostaFilho2017}.

Putting all of these elements together, we can briefly describe the operation of the experiment through our four step simulation process~[Fig.~\ref{fig:ExpSteps}].
During the first step, the double-well potential is added to all the ions. The coupling to the bath relaxes the ions.
During the second step, the double-well potential is removed from the active ions (B), while kept at the memory ones (A). The location of the memory ions is tracked through fluorescence imaging and the minimum of the harmonic potential is controlled so as to capture the PCA transition rule.
During the third step, the double-well potential is again added to all the ions.
During the fourth step, the second step is repeated exchanging the role of the A and B oscillators to be the active and the memory ions, respectively.

\vspace{5mm}
{\bf Superconducting circuits}\\

We now consider a superconducting circuit array as a system of tunable non-linear coupled oscillators.
We consider a flux-qubit architecture, whereby the oscillator ``position'' corresponds to the phase $\Phi$  generated by a persistent current  within the superconducting loop~\cite{orlando:1999}; physically, this persistent current enforces the requirement that exactly a multiple of the flux quantum $\Phi_0$ penetrates the superconducting loop.
In this configuration, the potential landscape can be tuned by changing the threaded flux through different elements of the superconducting qubit~\cite{quintana:2017}; this ability to straightforwardly engineer double-well potentials, immediately satisfies requirement {\bf (i)}.

The most direct way of coupling two superconducting oscillators is by proximatizing them and using their mutual conductance ($C_{\text{mut}}$) or inductance ($M_{\text{mut}}$)~\cite{krantz:2019}.
In our case, we are interested in leveraging inductance coupling which directly couples the phases between the oscillators $H_{I} = M_{\text{mut}} \sin \Phi_i \sin\Phi_j$ in analogy to the position dependent coupling discussed in the main text.
While proximitization enables the coupling between two oscillators, this coupling can also be mediated by an intermediate superconducting loop, allowing for additional tuning control of the interactions~\cite{krantz:2019}.

In general, implementing an arbitrary PCA will requires $N$-body interactions to capture the correct potential necessary in steps (2) and (4), where $N$ is the number of cells associated with the transition rule (4 in the case of the $\pi$-Toom model).
Fortunately, there has been extensive effort creating such interactions using superconducting qubits, which we can leverage when considering the continuum (classical) limit of these objects~\cite{menke:2022}.
Such interactions arise when multiple superconducting oscillators are inductively coupled to a tunable, common oscillator that is far detuned---this satisfies requirement {\bf (ii)}.
Let us note that, the additional connectivity, as well as the tuning of a larger number of interactions implies that the complexity of the object increases rapidly with the complexity of the PCA rule.

Curiously, the particular case of the $\pi$-Toom model can actually be implemented using only tunable two-oscillator interactions.
Given the anti-majority vote nature of the $\pi$-Toom transition, the potential:
\begin{align}
  V_{I,\text{approx}} &= \left[\frac{\Phi_{x,y}^A + \Phi_{x+1,y}^A + \Phi_{x,y+1}^A}{3} + \Phi_{x,y}^B \right]^2 \notag\\
  &= \frac{(\Phi_{x,y}^A + \Phi_{x+1,y}^A + \Phi_{x,y+1}^A)^2}{9} + (\Phi_{x,y}^B)^2 +  \frac{2}{3}\Phi^B_{x,y}\left[ \Phi_{x,y}^A + \Phi_{x+1,y}^A + \Phi_{x,y+1}^A\right] \label{eq:simple}
\end{align}
has a minimum where the sign of $\Phi_{x,y}^B$ is opposite to the mean of the nearby memory cells. Fixing the A particles as the memories using the pinning potential, the B oscillator can equilibrate to the potential minimum, implementing the necessary anti-majority transition rule.
Note that different values of the mean of the $\Phi^A$ phases will place the minimum of the potential of $\Phi^B_{x,y}$ at different locations; however, since the sign of the phase is correctly captured, we expect that the following pinning step causes the system to relax to the same phase value regardless of the differences in $\Phi^A_i$.

Having discussed how interactions can be generated to implement the PCA transition rules, we turn to discussing the role of the bath in the platform.
Much like the trapped ion case, one can directly control the temperature of the environment (e.g. by adjusting the cryostat temperature) in order to modify the bath coupled to the superconducting circuit array~\cite{quintana:2017}.
One can also artificially inject noise into the system and study how the ability to simulate the PCA dynamics is affected by the spectral properties of the injected noise~\cite{Sung19}.
These considerations satisfy requirement {\bf (iii)}.

Zooming out a bit, we note that this setup provides a flexible platform that can investigate broad questions related to the classical to quantum crossover in the dynamics of open systems.
% more than the protocol presented in our work.
%
Indeed, by lowering the temperature and preparing smaller  coherent states, one eventually approaches the regime where the quantum nature of the dynamics becomes important.
%
%What features would become manifest in this crossover remains unclear to us at this stage, and we leave a more in-depth analysis for future work.

\bibliography{myref}